\documentclass[preprint,nohyper]{JHEP3} %
\usepackage{epsfig,multicol}
\usepackage{latexsym}

\newcommand{\newc}{\newcommand}
\newc\eg{{\it {e.g.}}}  \newc\etal{{\it {et al.}}} \newc\ie{{\it i.e.}}
\newc\etc{{\it {etc}}}  

\newcommand\lsim{\mathrel{\rlap{\lower4pt\hbox{\hskip1pt$\sim$}}
    \raise1pt\hbox{$<$}}}
\newcommand\gsim{\mathrel{\rlap{\lower4pt\hbox{\hskip1pt$\sim$}}
    \raise1pt\hbox{$>$}}}

\newc{\mhalf}{m_{1/2}}      \newc{\mzero}{m_0}

\newc{\tanb}{\tan\beta}
\newc{\azero}{A_0}
\newc{\at}{A_t} \newc{\abot}{A_b} \newc{\atau}{A_\tau} 

\newc{\bmu}{B\mu}           \newc{\sgn}{{\rm sgn}}
\newc{\mone}{M_1}           \newc{\mtwo}{M_2}

\newc{\charone}{\chi_1^\pm} \newc{\mcharone}{m_{\chi_1^\pm}}

\newc{\hl}{h}               \newc{\mhl}{m_{\hl}}
\newc{\hh}{H}               \newc{\mhh}{m_{\hh}}
\newc{\ha}{A}               \newc{\mha}{m_{\ha}}
\newc{\hc}{H^{\pm}}         \newc{\mhc}{m_{\hc}}

\newc{\mgut}{M_{\rm GUT}}

\newc{\mplanck}{M_{\rm P}}      \newc{\mpl}{M_{\rm Pl}}
\newc{\msusy}{M_{\rm SUSY}}      \newc{\ms}{M_{\rm S}}

\newc{\jxf}{J({\xf})}
\newc{\jxfexact}{J_{\rm exact}({\xf})}  \newc{\jxfexp}{J_{\rm exp}({\xf})}
\newc{\VEV}[1]{\langle #1 \rangle}

\newc{\xf}{x_f}
\newc\vrel{v_{\rm rel}}
\newcommand\mchi{m_{\chi}}              

\newc\sell{{\widetilde e}_L}      \newc\msell{m_{\sell}}
\newc\selr{{\widetilde e}_R}      \newc\mselr{m_{\selr}}

\newc\snue{{\widetilde \nu}_e}      \newc\msnue{m_{\snue}}
\newc\snutau{{\widetilde \nu}_\tau}      \newc\msnutau{m_{\snutau}}

\newc\supl{{\widetilde u}_L}      \newc\msupl{m_{\supl}}
\newc\supr{{\widetilde u}_R}      \newc\msupr{m_{\supr}}
\newc\sdl{{\widetilde d}_L}      \newc\msdl{m_{\sdl}}
\newc\sdr{{\widetilde d}_R}      \newc\msdr{m_{\sdr}}

\newcommand{\stauone}{{\tilde \tau}_1}   \newcommand\mstauone{m_{\stauone}}

\newcommand{\stauright}{{\tilde \tau}_R}

\newcommand\gluino{\tilde g}
\newcommand\mgluino{m_{\gluino}}

\newc\hpm{H^\pm} \newc\hp{H^+} \newc\hm{H^-} 
\newc\sfermion{\tilde f}  \newc\msfermion{m_{\sfermion}}  

\newc\second{{\rm sec}} 

\newc\alphas{\alpha_s}

\newc\alphaem{\alpha_{em}}

\newcommand\mz{m_{Z}}

\newcommand\treh{T_{\rm R}}

\newc{\sthw}{\sin\theta_W}              \newc{\cthw}{\cos\theta_W}
\newc{\bino}{\widetilde B}              \newc{\wino}{\widetilde W_3}
\newc{\higgsinob}{{\widetilde H}^0_b}   \newc{\higgsinot}{{\widetilde H}^0_t}

\newc{\abund}{\Omega h^2}
\newc{\abundchi}{\Omega_\chi h^2}
\newc{\abundcdm}{\Omega_{{\rm CDM}} h^2}

\newc{\omegam}{\Omega_{{\rm M}}}       \newc{\abundm}{\Omega_{{\rm M}} h^2}
\newc{\omegab}{\Omega_{{\rm b}}}	\newc{\abundb}{\Omega_{{\rm b}} h^2}
\newc{\omegatot}{\Omega_{{\rm TOT}}}

\newc{\omeganlsp}{\Omega_{{\rm NLSP}}}   \newc{\abundnlsp}{\Omega_{\rm NLSP}h^2}     
\newc{\ynlsp}{Y_{{\rm NLSP}}}            \newc{\taunlsp}{\tau_{{\rm NLSP}}}
\newc{\nnlsp}{n_{{\rm NLSP}}}            \newc{\mnlsp}{m_{{\rm NLSP}}}
\newc{\nx}{n_{X}}                        \newc{\yx}{Y_{X}}
\newc{\mx}{m_{X}}                        \newc{\taux}{\tau_{X}}

\newc{\rhocrit}{\rho_{crit}}
\newc{\rhochi}{\rho_{\chi}}

\newcommand\fa{f_{a}}

\newcommand\stau{\tilde{\tau}}

\newcommand\neut{\tilde \chi}

\newc{\cachigamma}{C_{a\neut\gamma}}
\newc{\caww}{C_{aWW}}                   
\newc{\cayy}{C_{aYY}}

\newc{\nl}{\cos \theta_{\tilde t}}
\newc{\nr}{\sin \theta_{\tilde t}}
\newcommand\tev{\,\mbox{TeV}}
\newcommand\gev{\,\mbox{GeV}}
\newcommand\mev{\,\mbox{MeV}}
\newcommand\kev{\,\mbox{keV}}

\newc\gbar{{\overline{g}}}

\newc{\ra}{\rightarrow}

\newc{\beq}{\begin{equation}}
\newc{\eeq}{\end{equation}}
\newc{\bea}{\begin{eqnarray}}
\newc{\eea}{\end{eqnarray}}

\newc{\nspin}{n_{\rm spin}}
\newc{\nflavor}{n_{\rm F}}
\newc{\ngamma}{n_\gamma}
\newc{\ychi}{Y_{\chi}}                  \newc{\yeqchi}{Y^{\rm EQ}_{\chi}}

\newcommand\axino{\tilde{a}}

\newc{\naxino}{n_{\axino}}
\newc{\yaxino}{Y_{\axino}}
\newc{\yeqaxino}{Y^{\rm EQ}_{\axino}}
\newc{\ythaxino}{Y^{\rm TP}_{\axino}}
\newc{\ynthaxino}{Y^{\rm NTP}_{\axino}}

\newcommand\gravitino{\widetilde{G}}    
\newcommand\mgravitino{m_{\gravitino}}

\newcommand\abundg{\Omega_{\gravitino}h^2}
\newcommand\abundgntp{\Omega^{\rm NTP}_{\gravitino}h^2}     

\newcommand\abundgtp{\Omega^{\rm TP}_{\gravitino}h^2}       

\newc{\ngravitino}{n_{\gravitino}}
\newc{\ygravitino}{Y_{\gravitino}}
\newc{\yeqgravitino}{Y^{\rm EQ}_{\gravitino}}
\newc{\ythgravitino}{Y^{\rm TP}_{\gravitino}}
\newc{\ynthgravitino}{Y^{\rm NTP}_{\gravitino}}

\newc{\yascat}{Y^{\rm scat}_{i,j}}      \newc{\yadec}{Y^{\rm dec}_{i}}
\newc{\gstar}{g_\ast}           \newc{\gsstar}{g_{s\ast}}

       \def\pslash{\not{\hbox{\kern-2.3pt $p$}}}
       \def\kslash{\not{\hbox{\kern-2.3pt $k$}}}
       \def\qslash{\not{\hbox{\kern-2.3pt $q$}}}
       \def\ddslash{\not{\hbox{\kern-2.3pt $d$}}}
       \def\prtslash{\not{\hbox{\kern-2.3pt $\partial$}}}

\newcommand\jcap[3] 
		{{\it J.\ Cosmol.\ Astrop.\ Phys.\ }{\bf #1} (#2) #3}

\title{Gravitino Dark Matter in the CMSSM\\
  and Implications for Leptogenesis and the LHC}
\author{Leszek Roszkowski\\
Department of Physics and Astronomy,
University of Sheffield, Sheffield, S3 7RH, England\\
E-mail: \email{L.Roszkowski@sheffield.ac.uk}
}
\author{Roberto Ruiz de Austri\footnote{Present address: 
Departamento de F\'{\i}sica Te\'{o}rica C-XI
 and Instituto de F\'{\i}sica Te\'{o}rica C-XVI,
 Universidad Aut\'{o}noma de Madrid, Cantoblanco,
 28049 Madrid, Spain}\\
Department of Physics and Astronomy,
University of Sheffield, Sheffield, S3 7RH, England\\
E-mail: \email{R.RuizDeAustri@sheffield.ac.uk}
}
\author{Ki-Young Choi\\
Department of Physics and Astronomy,
University of Sheffield, Sheffield, S3 7RH, England\\
E-mail: \email{K.Choi@sheffield.ac.uk}
}

\abstract{
In the framework of the CMSSM we study the gravitino as the
lightest supersymmetric particle and the dominant
component of cold dark matter in the Universe. We include both a
thermal contribution to its relic abundance from scatterings in the
plasma and a non--thermal one from neutralino or stau decays after
freeze--out. In
general both contributions can be important,
although in different regions of the parameter space.
We further include constraints from BBN on electromagnetic and
hadronic showers, from the CMB blackbody spectrum and from collider
and non--collider SUSY searches. The region where the neutralino is
the next--to--lightest superpartner is severely constrained by a
conservative bound from excessive electromagnetic showers and probably
basically excluded by the bound from hadronic showers, while the
stau case remains mostly allowed. In both regions the constraint from
CMB is often important or even dominant.  In the stau case, for the
assumed reasonable ranges of soft SUSY breaking parameters, we find
regions where the gravitino abundance is
in agreement with the range inferred from CMB studies, provided that,
in many cases, a reheating temperature $\treh$ is large,
$\treh\sim10^{9}\gev$.  On the other side, we find an upper bound
$\treh\lsim 5\times 10^{9}\gev$. Less conservative bounds from BBN
or an improvement in measuring the CMB spectrum would provide a
dramatic squeeze on the whole scenario, in particular it would
strongly disfavor the largest values of $\treh\sim 10^{9}\gev$. The
regions favored by the gravitino dark matter scenario are very
different from standard regions corresponding to the neutralino
dark matter, and will be partly probed at the LHC.
}
 
\begin{document}

\section{Introduction}\label{sect:intro}

Low--energy supersymmetry (SUSY) provides perhaps the most attractive
candidates for cold dark matter (CDM) in the Universe. This is because
in the SUSY spectrum several new massive particles appear, some
of which carry neither electric nor color charges. The lightest among
them (the lightest SUSY partner, or the LSP) can then be neutral and either
absolutely stable by virtue of some discrete symmetry, like
$R$--parity, or very long--lived, much longer than the age of the
Universe, and thus effectively stable. A particularly well--known and
attractive example of such a weakly--interacting massive particle
(WIMP) is the lightest neutralino.  In recent years, however, there
has been also a renewed interest in the two alternative
well--motivated SUSY candidates for WIMPs and CDM, namely the
gravitino and the axino.

The spin--$3/2$ gravitino acquires its mass from spontaneous breaking
of local SUSY, or supergravity. Since its interactions with ordinary
matter are typically strongly suppressed by an inverse square of the
(reduced) Planck mass, cosmological constraints become an
issue~\cite{dz81}. Early on it was thought that, with a primordial
population of gravitinos decoupling very early, if stable, they had to
very light, below some $\lsim1\kev$~\cite{pp82}, in order not to
overclose the Universe, or otherwise very heavy,
$\gsim10\tev$~\cite{weinberg-grav82}, so that they could decay before
the period of BB nucleosynthesis (BBN). With inflation these bounds
disappear~\cite{khlopov+line84,ekn84} but other problems emerge when
gravitinos are re--generated after reheating. If the gravitino is not
the LSP, it decays late ($\sim10^8\sec$) into the LSP (say the
neutralino) and an energetic photon which can distort the abundances
of light elements produced during BBN, for which there is a good
agreement of calculations with direct observations and with CMB
determinations. Since the number density of gravitinos is directly
proportional to the reheating temperature $\treh$, this leads to an
upper bound of
$\treh<10^{6-8}\gev$~\cite{khlopov+line84,ekn84,ens84,nos83,jss85,km94}
(for recent updates see, \eg,~\cite{cefo02,kkm04}). On the other hand,
when the gravitino is the LSP and stable, ordinary sparticles can
decay into it and an energetic photon. A combination of this and the
overclosure argument ($\abundg<1$) leads in this case to an upper
bound $\treh\lsim 10^9\gev$~\cite{ens84,mmy93}.

Because of many similarities, for comparison we comment here on axinos
as DM.  The axino is a fermionic superpartner of the axion in
supersymmetric models with the Peccei--Quinn (PQ) mechanism
implemented for solving the strong CP problem~\cite{pq}. Axino
interactions with ordinary matter are suppressed by $1/\fa^2$, where
$\fa\sim10^{11}\gev$ is the PQ scale.  In many models the axino mass is
not directly determined by a SUSY breaking scale $\msusy$, in contrast
to the neutralino and the gravitino, and therefore the axino can naturally be the
LSP. Without inflation the axino has to be light ($\lsim1\kev$) and
thus warm DM~\cite{kmn,bgm,rtw,ckkr,ay00}. Otherwise it can naturally
be a cold DM~\cite{ckr,ckkr,crs1}, so long as $\treh\lsim
10^{4-5}\gev$.  Constraints from Big Bang Nucleosynthesis (BBN) on
axino CDM are relatively weak since NLSP decays to them typically take
place before BBN.  See~\cite{bs04} for an improved treatment of
thermal production of axinos at $\treh\gsim 10^{5}\gev$.

For both the gravitinos and the axinos, despite their typical
interaction strengths being so much weaker than electroweak, their
relic abundance can still be of the favored value of $\sim0.1$.  This is
because they can be efficiently produced in a class of {\em thermal
production} (TP) processes involving scatterings and decays of
particles in the primordial plasma, depending on
$\treh$. Alternatively, in a {\em non--thermal production} (NTP) class
of processes, the next--to--lightest supersymmetric particle (NLSP)
first freezes out and next decays to the axino or gravitino.  These
mechanisms are supposed to re--generate the relics, after their
primordial population has been diluted by a proceeding period of
inflation.

In TP, gravitino (or, alternatively, axino) production proceeds
predominantly through ten classes of processes involving
gluinos~\cite{ekn84,mmy93}. In four of them, a logarithmic singularity
appears due to a $t$--channel exchange of a massless gluon which can
be regularized by introducing a thermal gluon
mass~\cite{ekn84,mmy93}. A full result for the singular part was
obtained in~\cite{bbp98}. In~\cite{bbb00} a resumed gluon propagator
was used to obtain the finite part of the production rate, and an
updated expression for the relic abundance $\abundgtp$ of gravitinos
generated via TP, valid at high $\treh$, was given
\begin{equation}
\abundgtp\simeq 0.2 \left(\frac{\treh}{10^{10}\gev}\right)
\left(\frac{100\gev}{\mgravitino}\right) 
\left(\frac{\mgluino(\mu)}{1\tev}\right)^2,
\label{eq:abundgbbb}
\end{equation}
where $\mgluino(\mu)$ above is the running gluino
mass. In~\cite{bbp98,bbb00} it was argued that, for natural ranges of
the gluino and the gravitino masses, one can have $\abundgtp\sim 0.1$
at $\treh$ as high as $10^{9-10}\gev$.  Such high values of
$\treh$ are essential for thermal leptogenesis~\cite{fy,crv96}, with a
lower limit of $\treh>2\times 10^9\gev$~\cite{gnrrs04}.

The issue of gravitino relics generated in NTP processes and
associated constraints  was recently re--examined in
detail in
~\cite{feng03-prl,feng03,fst04-slep,fst04-sugra} and
in~\cite{eoss03-grav}. 
Since all the NLSPs decay into gravitinos, in this case
\beq
\abundgntp=\frac{\mgravitino}{\mnlsp}\Omega_{\rm NLSP} h^2,
\label{abundgntp:eq}
\eeq
where $\abundgntp$ is the NTP contribution to the gravitino relic
abundance and $\abundnlsp$ would have been the relic abundance of the NLSP if it
had remained stable. Note that $\abundgntp$ grows with $\mgravitino$. 

In grand--unified SUSY frameworks, like the popular Constrained
Minimal Supersymmetric Standard Model (CMSSM)~\cite{kkrw94}, which
encompasses a class of unified models where at the GUT scale gaugino soft
masses unify to $\mhalf$ and scalar ones unify to $\mzero$, one often
finds that the relic abundance of the neutralino (or the stau) is
actually significantly larger than $0.1$. One therefore in general
cannot neglect the 
contribution from the NTP mechanism, unless
$\mgravitino$ is rather small. In this case, however, at high
$\treh\sim10^{9}\gev$, TP contribution is likely to play a
role, since $\abundgtp\propto{\treh/\mgravitino}$. Clearly, in general
both TP and NTP must be simultaneously considered.

In~\cite{fst04-sugra} NTP of gravitinos was considered in an effective
low--energy SUSY scenario.  The relic abundance of gravitinos from NTP
via neutralino, stau (which had already been examined in gauge--mediated
SUSY breaking schemes in~\cite{ahs00}) and sneutrino NLSP decays was
however only crudely approximated and that from TP was not included
at all. A
weak constraint $\abundgntp\lsim0.1$ (rather than $\abundgntp\sim0.1$)
was assumed. On the other hand, 
constraints from BBN were treated with much care. Typically, lifetimes
for NLSP decays into gravitinos are $\sim10^8\sec$ in which case
constraints from electromagnetic fluxes are particularly
important. Nevertheless, hadronic showers, which in the past were thought to be 
important only for lifetimes $\lsim10^4\sec$, must also be included in
considering particle decays in late times, since they provide
additional strong constraints. In the case of light
gravitinos in gauge--mediated SUSY breaking schemes this constraint
was applied in~\cite{ggr98} and in the case of CDM gravitinos
in~\cite{fst04-slep,fst04-sugra}.  Furthermore,
in~\cite{fst04-slep,fst04-sugra} substantial constraints on the SUSY
parameter space were derived from a requirement of not distorting the
CMB blackbody spectrum by energetic photons~\cite{dz81,ekn84}.

NTP of gravitinos in the framework of the CMSSM was examined
in~\cite{eoss03-grav}. Constraints from electromagnetic showers were
applied, but not from hadronic ones. Nor was the constraint from CMB
applied.  Gravitino abundance from NTP was computed much more
accurately than
in~\cite{feng03-prl,feng03,fst04-slep,fst04-sugra}. On the other hand,
similarly to~\cite{feng03-prl,feng03,fst04-slep,fst04-sugra}, only
cosmologically allowed regions $\abundgntp\lsim0.1$ were 
delineated but not cosmologically favored ones of
$\abundgntp\simeq0.1$. For the assumed ranges of parameters some
regions were found where $\abundgntp$ was not excessively large, but
actually too low. (In~\cite{eoss03-grav} it was also noted that any
possible stau NLSP asymmetry would be washed away by stau
pair--annihilation into tau pairs.) 
As we will show later, unlike the authors of~\cite{eoss03-grav}, in
the CMSSM at large values of $\mhalf$ (beyond those considered
in~\cite{eoss03-grav}) we have found cosmologically favored regions of
$\abundgntp\simeq0.1$.

A question arises for which values of SUSY parameters, as well as
$\treh$, the combination of gravitino yields from both TP and NTP
gives $\abundg\sim0.1$ in unified SUSY schemes.  In this paper we
investigate this issue within the CMSSM, which is a model of much
interest. We assume no specific underlying supergravity model and
treat $\mgravitino$ as a free parameter. In the CMSSM soft masses are
assumed to be generated via a gravity--mediated SUSY breaking
mechanism, in which case $\mhalf$, $\mzero$ and $\mgravitino$ can be
in the $\gev$ to $\tev$ range, and we need to ensure that the
gravitino is the LSP.  We compute the relic abundance of gravitinos
with high accuracy which matches present observational precision of
CDM abundance determinations. In evaluating $\abundgtp$ we
follow~\cite{bbb00}, while $\abundgntp$ is determined by the yield of
the NLSP which we compute numerically, following our own calculation
(without partial wave expansion), as described below.  We apply
constraints from both the electromagnetic and from the hadronic
fluxes and from CMB spectrum, as well as the usual constraints from
collider and non--collider SUSY searches. We concentrate on the
largest $\treh\sim10^{9}\gev$ but also consider the region of
low $\treh$ where NTP dominates.

Before proceeding, we should note that there are other possible
gravitino production mechanisms, \eg\ via inflaton decay or during
preheating~\cite{grt99,kklp99}, but they are much more model dependent
and not necessarily efficient~\cite{nps01}. Alternatively, gravitinos
may be produced from decays of moduli fields~\cite{kyy04}. In this
paper, we do not include these effects.

We further assume $R$--parity conservation, both for simplicity and
because otherwise it is hard to understand why weak universality works
so well.  However, it is worth remembering that in the case of such
super--weakly interacting relics as the axino or the gravitino,
$R$--parity is not really mandatory, unlike in the case of the
neutralino WIMP. Indeed, the suppression provided by the PQ or Planck
scale is often sufficient to ensure effective stability of such relics
on cosmological time scales even when $R$--parity breaking terms are
close to their present upper bounds. Indeed, in the case of the axino
CDM, a tiny amount of its decay products into $e^+e^-$ pairs has been
proposed as an interesting way of explaining an apparent INTEGRAL
anomaly~\cite{hw04}.

In the following, we will first summarize our procedures for computing
$\abundg$ via both TP and NTP. Then we will list NLSP decay modes into
gravitinos, and discuss constraints on the CMSSM parameter space, in
particular those from BBN and CMB. Finally, we will discuss
implications of our results for thermal leptogenesis and for SUSY
searches at the LHC.

\section{Gravitino Relic Abundance}\label{sec:abundg}

The present relic abundance of any stable, massive relics ($\chi$,
$\stau$, $\axino$, $\gravitino$, \ldots) produced either thermally or
non-thermally is related to their yield\footnote{We define the yield
as $Y=n/s$, where $s=(2\pi^2/45)g_{\ast s}T^3$ is the entropy density,
following a common convention of~\cite{kt90} which is also used
in~\cite{kkm04}. Another definition, used, \eg,
in~\cite{fst04-sugra,cefo02}, is 
$Y^\prime=n/n_\gamma$ 
where
$n_\gamma$ is the number density of photons in the CMB, $n_\gamma=2
n_{rad}=2\zeta(3)T^3/\pi^2$. At late times $t\gsim10^6\sec$, typical
for NLSP decays to gravitinos, and later, $s\simeq 7.04 n_\gamma$.}
as
\begin{equation}
Y=3.7\times 10^{-12} \left(\frac{100\gev}{m}\right) 
\left(\frac{\Omega h^2}{0.1}\right).
\label{yoh2:eq}
\end{equation}
where $m$ is the mass of the relic particle.

\subsection{TP}\label{sec:tp}

The yield of massive relics generated through TP processes can be obtained
by integrating the Boltzmann equation with both scatterings and decays
of particles in the expanding plasma~\cite{kt90}. In the case of the
gravitino LSP, dominant contributions come from 2--body
processes involving
gluinos~\cite{ekn84,mmy93,bbp98,bbb00}. 
These are given by a dimension--5 part of the Lagrangian describing
gravitino interactions with gauge bosons and
gauginos.\footnote{See, \eg,~\cite{moroi95}.}
For the ten classes of scattering
processes the cross section, at large energies, has the form
\begin{equation}
\sigma(s) \propto \frac{1}{\mplanck^2}\left(1 +
\frac{\mgluino^2}{3\mgravitino^2}\right),~~~~~~~~~~~~~s\gg\msusy,
\label{tpcs:eq}
\end{equation}
where $\mplanck=1/\sqrt{8\pi G_N}=2.4\times10^{18}\gev$ is the reduced
Planck mass.

In actual computations we solve the Boltzmann equation numerically by
following the usual steps described, \eg, in~\cite{bbb00,ckkr}, and
use the expression~(44) of~\cite{bbb00} for the sum of soft and hard
contributions to the collision terms.

We do not include gluino decays into gravitinos which, like in the
case of axino LSP, would only become important at
$\treh\sim\mgluino$~\cite{mmy93,ckkr}. This is because we concentrate on
large $\treh\sim10^{9}\gev$ which are relevant for models of thermal
leptogenesis.

\subsection{NTP}\label{sec:ntp}

In computing the relic abundance of gravitinos generated through NTP
processes we first compute their yield after freezeout. In the CMSSM
in most cases the NLSP is either the (bino--like) neutralino or the
lighter stau.  In the case of the neutralino, we include exact cross
sections for all the tree--level two--body neutralino processes of
pair--annihilation~\cite{nrr1,nrr2} and coannihilation with the
charginos, next--to--lightest neutralinos~\cite{gondoloedsjo} and
sleptons~\cite{nrr3}.  This allows us to accurate compute the yield
and $\abundchi$ in the usual case when the lightest neutralino is the
LSP.  We further extend the above procedure to the case where the NLSP
is the lightest stau $\stauone$ (a lower mass eigenstate of $\stau_R$
and $\stau_L$). We include all slepton--slepton annihilation and
slepton--neutralino coannihilation processes.  In both cases we
numerically solve the Boltzmann equation for the NLSP yield and use
exact (co)annihilation cross sections which properly take into account
resonance and new final--state threshold effects.  The procedure has
been described in detail in~\cite{rrn1} and was recently applied to
the case of axino LSP~\cite{crrs}.

\section{NLSP Decays into Gravitinos}\label{sec:nlspdecays}

Once the NLSPs freeze out from thermal plasma at $t\sim10^{-12}\sec$,
their comoving number density remains basically constant until they start
decaying into gravitinos at late times $t\sim10^{4}-10^{8}\sec$ (so
long as $\mgravitino\gsim0.1\msusy$ which we assume here). Associated
decay products generate energetic fluxes which are mostly
electromagnetic (EM) but also hadronic (HAD). If too large, these will
wreck havoc on the abundances of light elements.  Limits on
electromagnetic showers become important for $t\gsim10^{4}\sec$ and
come mainly either from excessive deuterium destruction via
$\gamma+D\ra n+p$ ($10^{4}\sec\lsim t\lsim10^{6}\sec$), or production
via $\gamma+{^4}{\! He}\ra D+\ldots$ ($t\gsim10^{6}\sec$).  Decay rates and
branching ratios into EM radiation generated by late--decaying
particles have recently been re--analyzed by Cyburt,
\etal, in~\cite{cefo02}. Updated bounds on EM fluxes have been obtained
on the parameter $\zeta_X= m_X n_X/n_\gamma$, where $X$ denotes the
decaying particle,\footnote{Hereafter we will mean $X=\chi,\stauone$
for brevity.} $n_\gamma$ is a number density of background photons, as
a function of $X$ lifetime $\tau_X$, by assuming that in
$X\ra\gravitino+\ldots$ decays associated showers are mostly
electromagnetic. This is indeed the case when $X$ is either the
neutralino or the stau~\cite{eoss03-grav,fst04-slep}. 

Limits from BBN on hadronic
showers are stronger for
$\tau\lsim10^4\sec$ but are often also important at later
times~\cite{renoseckel88,dehs88+89,kkm04,jedamzik04}. 
They come mainly from ${^4}{\! He}$ overproduction via $n+p\ra D\ra {^4}{\! He}$
for $\taux\lsim10^2\sec$, and $D$ overproduction via $n+{^4}{\! He}\ra
D$ at later times.  Hadronic components produced in late $X$ decays
into gravitinos, while much less frequent, will still lead to
important constraints on the parameter space, as mentioned above, and
will play an important role in our analysis. Upper limits on hadronic
radiation from $X$ decays, but with somewhat stronger assumptions than
in~\cite{cefo02}, have recently been re--evaluated by Kawasaki, \etal,
in~\cite{kkm04}.

More stringent bounds, by roughly a factor of ten, come from
considering constraints from ${^6}{\! Li}$~\cite{jedamzik99} and/or
from ${^3}{\!  He}$~\cite{sjsb95}. As discussed, \eg,
in~\cite{fst04-slep} at present both are probably still too poorly
determined to be treated as robust. For this reason, and in order to
remain conservative, we do not use the constraint from ${^6}{\! Li}$,
unlike in~\cite{eoss03-grav}. Nor, like in~\cite{eoss03-grav}, do we
apply the constraint from ${^3}{\!  He}$.

In order to apply bounds from BBN light element abundances on EM/HAD
showers produced in association with gravitinos, we need to evaluate
the relative energy $\xi^X_{i}$ ($i={em},\, {had}$), as defined below,
which is released in NLSP decays into EM/HAD radiation.
First, for each NLSP
decay channel we need to know the energy $\epsilon^X_{i}$
transferred to EM/HAD fluxes.  In decays $X\ra\gravitino+ R+\ldots$,
where $R$ collectively stands for all the particles generating either
EM or HAD radiation, the total energy per NLSP decay carried by $R$
will be a fraction of $\mx$. This is because, at late times of
relevance to $\gravitino$ production, the NLSPs decay basically at
rest.  In 2--body decays
\begin{equation}
E^X_{tot}= \frac{\mx^2-\mgravitino^2+m_R^2}{2 \mx},
\label{etotaldef:eq}
\end{equation}
where now $m_R$ stands for mass of $R$.  Unless $\mgravitino$ is not
much less than $\mx$, then, for negligible $m_R$, the usual
approximation $E^X_{tot}\simeq \mx/2$ works well. In the case of 3--
and more--body final states $E^X_{tot}$ as a fraction of $\mx$ assumes
a range of values.

We also need to compute the NLSP lifetime $\taux$ and branching
fractions $B^X_{i}$ ($i={em},\, {had}$) into EM/HAD showers.  All the above
quantities depend on the NLSP and (with the exception of the yield) on
its decay modes and the gravitino mass. For the cases of interest
($\chi$ and $\stauone$) these have been recently evaluated in detail
in~\cite{fst04-slep,fst04-sugra} (see also~\cite{eoss03-grav}) and
below we follow their discussion.

For the neutralino NLSP the dominant decay mode is
$\chi\ra\gravitino\gamma $ for which the decay rate
is~\cite{eoss03-grav,fst04-sugra} 
\begin{equation}
\Gamma\left(\chi\ra\gravitino\gamma
  \right)=\frac{|N_{11}\cos\theta_W+N_{12}\sin\theta_W |^2}
{48\pi\mplanck^2} \frac{\mchi^5}{\mgravitino^2} \left(1-
  \frac{\mgravitino^2}{\mchi^2}\right)^3 \left(1+
  3\frac{\mgravitino^2}{\mchi^2}\right),
\label{gamchitogammagino:eq}
\end{equation}
where $\chi=N_{11} \widetilde{B} + N_{12} \widetilde{W}^0_3 + N_{13}
\widetilde{H}^0_b + N_{14} \widetilde{H}^0_t$. In the CMSSM the
neutralino is a nearly pure bino, thus $\chi\simeq \widetilde{B}$.
The decay $\chi\ra\gravitino\gamma$ produces mostly EM energy.  Thus 
\begin{eqnarray}
B^\chi_{em} &\simeq & 1, \\
\epsilon^\chi_{em} & = & \frac{\mchi^2-\mgravitino^2}{2\mchi}
\label{brchiem:eq}
\end{eqnarray}
and the energy  $\xi^\chi_{em}$ released into
    electromagnetic showers is in this 
    case simply given by  
\begin{eqnarray}
\xi^\chi_{em}\simeq \epsilon^\chi_{em} B^\chi_{em} Y^\chi.
\label{xichiem:eq}
\end{eqnarray}

If kinematically allowed, the neutralino can also decay via
$\chi\ra \gravitino Z, \gravitino h, \gravitino H, \gravitino A$ for
which the decay rates are given in~\cite{eoss03-grav,fst04-sugra}.
These processes contribute to hadronic fluxes because of large
hadronic branching ratios of the $Z$ and the Higgs bosons
($B^Z_{had}\simeq 0.7$, $B^h_{had}\simeq 0.9$). In this case  the energy
     $\xi^\chi_{had}$ released into 
    hadronic showers is
\begin{eqnarray}
\xi^\chi_{had}\simeq \left( \sum{h} \epsilon^\chi_{had} B^\chi_{had}\right)
Y^\chi
\label{xichihad:eq}
\end{eqnarray}
where the sum goes over all hadronic decay modes,
%
\begin{eqnarray}
\sum \epsilon^\chi_{had} B^\chi_{had}\simeq 
\frac{\epsilon^\chi_Z \Gamma(\chi\ra\gravitino Z)B^Z_{had} +
  \sum\epsilon^\chi_{h}\Gamma(\chi\ra\gravitino h)B^h_{had}  +
  \epsilon^\chi_{q\bar{q}}\Gamma(\chi\ra\gravitino q\bar q)}
{\Gamma\left(\chi\ra\gravitino\gamma \right) + 
 \Gamma\left(\chi\ra\gravitino Z \right) +
  \Gamma\left(\chi\ra\gravitino h \right)},
\label{brchihad:eq}
\end{eqnarray}
where 
\begin{eqnarray}
\epsilon^\chi_{k} &\approx &
\frac{\mchi^2-m_{\gravitino}^2+m_{k}^2}{2\mchi}, \qquad \textrm{for}
\ k=Z,h,H,A,\\ 
\epsilon^\chi_{q\bar{q}} &\approx & \frac{2}{3}(m_\chi-m_{\tilde{G}}).
\label{epschihad:eq}
\end{eqnarray}

Below the kinematic threshold for neutralino decays into $\gravitino$
and the $Z$/Higgs boson, one needs to include 3--body decays with the
off--shell photon or $Z$ decaying into quarks for which
$B^\chi_{had}(\chi\ra\gravitino \gamma^\ast/Z^\ast \ra \gravitino
q\bar q)\sim 10^{-3} $~\cite{fst04-sugra}. This provides a lower bound
on $B^\chi_{had}$.  At larger $\mchi$ Higgs boson final states become
open and we include neutralino decays to them as well.

The dominant decay mode of the stau $\stauone$ 
is $\stauone\ra\gravitino\tau$ for which (neglecting the tau--lepton
mass) the decay width is~\cite{eoss03-grav,fst04-sugra}
\begin{equation}
\Gamma\left(\stauone\ra\gravitino\tau
  \right)=\frac{1}{48\pi\mplanck^2} \frac{\mstauone^5}{\mgravitino^2}
  \left(1- \frac{\mgravitino^2}{\mstauone^2}\right)^4.
\label{gamstautotaugino:eq}
\end{equation}

In~\cite{feng03,fst04-slep,fst04-sugra} it was argued that decays of staus
contribute basically only to EM showers, despite the fact that
a sizable fraction of tau--leptons decay into light mesons, like pions and
kaons. These decay electromagnetically much faster than the typical
time scale of hadronic interactions, mainly because at such late times
there are very few nucleons left to interact with~\cite{feng03}. Thus
\begin{eqnarray}
B^{\stauone}_{em} &\simeq & 1, \\
\epsilon^{\stauone}_{em} &\approx &\frac{1}{2}
\frac{\mstauone^2-\mgravitino^2}{2\mstauone},
\label{brstauem:eq}
\end{eqnarray}
where the additional factor of $1/2$ appears because about half of the
energy carried by the tau--lepton is transmitted to final state
neutrinos. The energy  $\xi^{\stauone}_{em}$ released into
    electromagnetic showers is in this case
\begin{eqnarray}
\xi^{\stauone}_{em}\simeq \epsilon^{\stauone}_{em} B^{\stauone}_{em} Y^{\stauone}.
\label{xistauem:eq}
\end{eqnarray}

As shown in~\cite{fst04-slep}, for stau NLSP, the leading contribution
to hadronic showers come from 3--body decays
${\stauone}\ra\gravitino\tau Z, \gravitino\nu_{\tau}W$, or from 4--body
decays ${\stauone}\ra\gravitino\tau \gamma^\ast/Z^\ast \ra
\gravitino\tau q \bar q$. The
corresponding energy  $\xi^{\stauone}_{had}$ is
\begin{eqnarray}
\xi^{\stauone}_{had}\simeq \left( \sum \epsilon^{\stauone}_{had}
B^{\stauone}_{had}\right) 
Y^{\stauone}
\label{xistauhad:eq}
\end{eqnarray}
where the sum goes over all hadronic decay modes, 
\begin{eqnarray}
\sum \epsilon^{\stauone}_{had} B^{\stauone}_{had}\simeq
\frac{\epsilon^{\stauone}_{Z} \Gamma(\stauone\ra\gravitino \tau  Z)B^Z_{had} + 
\epsilon^{\stauone}_{W} \Gamma(\stauone\ra\gravitino \nu_\tau W)B^W_{had} +
\epsilon^{\stauone}_{q\bar{q}} \Gamma(\stauone\ra\gravitino \tau q\bar q )}
{\Gamma\left(\stauone\ra\gravitino\tau \right)},   
\label{brstauhad:eq}
\end{eqnarray}
and 
\begin{eqnarray}
\epsilon^{\stauone}_{Z}
\simeq\epsilon^{\stauone}_{W}\simeq\epsilon^{\stauone}_{q\bar{q}}
\approx \frac{1}{3}(m_{\stauone}-m_{\gravitino}). 
\label{epsstauhad:eq}
\end{eqnarray}
One typically finds~\cite{fst04-sugra} $B^{\stauone}_{had}\sim10^{-5}-
10^{-2}$ when 3--body decays are allowed and $\sim10^{-6}$ from from
4--body decays otherwise, thus providing a lower limit on the
quantity. Given such a large variation in $B^{\stauone}_{had}$, the
choice~(\ref{epsstauhad:eq}) is probably as good as any other.

\section{Constraints}\label{sec:const}

\paragraph{$\bullet$~~\underline{The relic abundance}}

We will be mostly interested in the cases where the sum
$\abundg=\abundgtp+\abundgntp$ satisfies the $2\,\sigma$ range for
non--baryonic CDM 
\begin{equation}
0.094< \abundg < 0.129,
\label{eq:abundgtot}
\end{equation}
which follows from combining WMAP results \cite{wmap_cdm} with other
recent measurements of the CMB. Larger values are excluded.  Lower
values are allowed but disfavored. We will also delineate regions
where $\abundgntp$ alone satisfies the
range~(\ref{eq:abundgtot}). These regions will be cosmologically
favored for $\treh\ll 10^{9}\gev$ when TP can be neglected.

\paragraph{$\bullet$~~\underline{Electromagnetic and hadronic showers
    and the BBN}} 

As stated above, bounds on electromagnetic fluxes have recently
been re--evaluated and significantly improved in~\cite{cefo02}, while on
hadronic ones in~\cite{kkm04}. (A clear summary of the leading
constraints can be found in~\cite{fst04-slep}.) As mentioned above,
each analysis uses somewhat different assumptions and input
parameters.

In constraining EM showers, Cyburt, \etal,~\cite{cefo02} 
imposed the following observational bounds on light element abundances
\begin{eqnarray}
\label{cefodeutinput:eq}
1.3\times 10^{-5} &< D/H < & 5.3 \times 10^{-5}, \\
\label{cefoypinput:eq}
0.227 & < Y_p < & 0.249,\\
9.0\times 10^{-11} &< \left({^7}{\! Li}/{H}\right) < & 2.8\times 10^{-10}.
\label{cefoliinput:eq}
\end{eqnarray}
In the ${^4}{\! He}$ abundance ($Y_p$) the error bars were taken at $2\,\sigma$.

In deriving constraints from EM showers, we impose
$\xi^{X}_{em}<\left(\zeta_{X}/2\right)(n_\gamma/s)$ where $\zeta_{X}$
is defined above and in~\cite{cefo02} and the factor
of $n_\gamma/s \simeq 1/7.04$ is due to a difference in our
definitions of the yield.  The largest allowed values of $\zeta_{X}$,
as a function of $\taux$, for the assumed abundances of
$D/H+Y_p+{^7}{\! Li}$ can be read out of fig.~7 of~\cite{cefo02}.  We
fix the baryon--to--entropy ratio $\eta$ at $6.0\times10^{-10}$, which
is consistent with the WMAP result $\eta = 6.1^{+0.3}_{-0.2} \times
10^{-10}$~\cite{wmap_eta}.


In constraining hadronic fluxes from NLSP decays we
compare with Kawasaki, \etal,~\cite{kkm04}, who used the following input
\begin{eqnarray}
\label{kkmdeutinput:eq}
2.0\times 10^{-5} &< D/H < & 3.6\times 10^{-5}, \\
0.228 & < Y_p < & 0.248.
\label{kkmypinput:eq}
\end{eqnarray}
More specifically, we impose $\xi^{X}_{had}< E_{vis}Y_X$
where $E_{vis}Y_X$ is defined in~\cite{kkm04}. 
The largest allowed values
of $E_{vis}Y_X$, as a function of
$\taux$, for the assumed abundances of 
$D/H~({\rm low}) +Y_p(IT)$ can be read out of fig.~43 of~\cite{kkm04}, for which 
$B_{had}=1, m_X=1\tev$ and the effect of photo--dissociation due to
electromagnetic showers was turned off. 

Note that in ~\cite{kkm04} much less conservative error bars on $D/H$
were used than in~\cite{cefo02}. 
Our constraints from HAD showers are
therefore going to be accordingly somewhat  more uncertain than from EM
ones, despite not applying the constraint from ${^7}{\! Li}$ at all. 

As we will see, the ranges of supersymmetric parameters excluded by
the above constraints will strongly depend on the assumed
abundances of the light elements. Below we will illustrate this point
explicitely in the case of the EM showers by considering the case
$\xi^{X}_{em}< E_{vis}Y_X$ where $E_{vis}Y_X$ in this case is read off
from fig.~42 of~\cite{kkm04} where $B_{had}=0$.

It would be helpful to have available in the literature bounds on the
hadronic fluxes assuming more generous inputs,
like~(\ref{cefodeutinput:eq})--(\ref{cefoliinput:eq}), and to allow
$B_{had}$ to vary. On the other hand, the dependence in figs.~38--43
of~\cite{kkm04} on $m_X$ is rather weak.

\paragraph{$\bullet$~~\underline{CMB}}
As originally pointed out in~\cite{ekn84} and recently re--emphasized
in~\cite{feng03-prl,feng03}, late injection of 
electromagnetic energy may distort the frequency dependence of the CMB
spectrum from its observed blackbody shape. At late times of
interest, energetic photons from
NLSP decays lose energy through such processes as $\gamma
e^-\rightarrow \gamma e^-$ but photon number remains conserved, since
other processes, like double Compton scatterings and thermal
bremsstrahlung, become inefficient. As a result, the spectrum follows
the Bose--Einstein distribution function
\begin{equation}
f_\gamma(E)=\frac{1}{e^{E/(kT)+\mu}-1},
\label{eq:cmbbedist}
\end{equation}
where $\mu$ here denotes the chemical potential. 
The current bound is~\cite{mubound}
\begin{equation}
|\mu| < 9\times10^{-5}.
\label{eq:mubound}
\end{equation}
At decay lifetimes of $\taux \lsim 4\times 10^{11}\,\Omega_b
h^2\,\sec\simeq 8.8\times10^{9} \sec$, where $\Omega_b h^2$ is the
abundance of baryons, the bound~(\ref{eq:mubound}) leads to a
constraint on the EM energy release $\xi^{X}_{em}$ defined previously
via the relation~\cite{hu93}
\begin{eqnarray}
\mu = 8.01\times 10^2\left[\frac{\taux}{1\sec}\right]^{1/2}
e^{-(\tau_{dC}/\taux)^{5/4}}\times
\left(\frac{7.04\,\xi^{X}_{em}}{1\gev}\right), 
\end{eqnarray}
where
%
\begin{eqnarray}
\tau_{dC}&=&1.46\times10^8(T_0/2.7\,K)^{-12/5}(\Omega_b
  h^2)^{4/5}(1-Y_p/2)^{4/5} \sec\\ &\simeq &  6.085\times10^6 \sec 
\end{eqnarray}
and we have taken 
$T_0=2.725\, K$, $\Omega_b h^2=0.022$ and
  $Y_p=0.24$. This leads to
\begin{eqnarray}
\xi^{X}_{em} <& 1.59\times 10^{-8}\,
  e^{(\tau_{dC}/\taux)^{5/4}}\left(\frac{1 \sec}{\taux} \right)^{1/2}.
\label{eq:muboundonxiemall}
\end{eqnarray}

At later decay lifetimes ($\taux \gsim 4\times 10^{11}\,\Omega_b
h^2\,\sec\simeq 8.8\times10^{9} \sec$), elastic 
Compton scatterings are not efficient enough to lead to the
Bose--Einstein spectrum~\cite{hu93}. Instead, the CMB spectrum can be
described by the Compton $y$ parameter, $4y=\delta\epsilon/\epsilon$
given by
\begin{eqnarray}
\frac{\delta\epsilon}{\epsilon}=7.04\times\frac{c^2}{kT(t_{\rm eff})}\xi_{em}^{X},
\label{eq:cmbypar}
\end{eqnarray}
where $T(t)$ is the CMB temperature and $t_{\rm
eff}=[\Gamma(1-\beta)]^{1/\beta}\taux$, with the Gamma function
$\Gamma$ for a time--temperature relation $T\propto t^{-\beta}$.
In the relativistic energy dominated era in the early Universe, for $T
< 0.1 \mev$, 
\begin{eqnarray}
T=1.15\times 10^{-3}\left(\frac{t}{1\sec}\right)^{-1/2} \gev, 
\end{eqnarray}
which gives
$\beta=1/2$.  Thus $t_{eff}= [\Gamma(1/2)]^2 \taux=\pi\taux$.

The
present upper limit on $y$ parameter is \cite{pdg02} 
\begin{eqnarray}
|y|<1.2\times 10^{-5}. 
\label{eq:ybound}
\end{eqnarray}
This translates into an upper bound on $\xi_{em}^{X}$
\begin{eqnarray}
\xi^{X}_{em} < 7.84\times 10^{-9}\left(\frac{\pi \taux}{1\sec}\right)^{-1/2}
\simeq 4.42\times 10^{-9}\frac{1}{\sqrt{\taux}}.
\label{eq:yboundonxiem}
\end{eqnarray}
%
Thus we can see that, at the late times $\taux \gsim8.8\times10^{9}
\sec$, as specified above, the constraint on the parameter space coming
from the $y$ parameter is applicable while at earlier times the
$\mu$--parameter constraint is applicable~\cite{hu93}.

\paragraph{$\bullet$~~\underline{Collider and Non--Collider Bounds}}
The relevant bounds from LEP are from chargino and Higgs
searches~\cite{lepbounds} 
\begin{eqnarray}
\label{eq:lepcharginobound}
\mcharone &>&104\gev,\\
m_h&>&114.4\gev.
\label{eq:lephiggsbound}
\end{eqnarray}
In addition, a good agreement of the measured $BR(B\rightarrow
X_s\gamma)$ with a Standard Model prediction places strong 
constraints on potential SUSY contributions to the process, which at
large Higgs VEV ratio $\tanb$ can be substantial. We impose~\cite{or1+2}
\begin{eqnarray}
BR(B\rightarrow
X_s\gamma) =  (3.34 \pm 0.68)\times10^{-4}.
\label{eq:bsgbounds}
\end{eqnarray}

\noindent
Finally, we exclude cases leading to tachyonic solutions and
those for which the gravitino is not the LSP.

\section{Results}\label{sec:results}

Mass spectra of the CMSSM are determined in terms of the usual five
free parameters: the previously mentioned $\tanb$, 
$\mhalf$ and $\mzero$, as well as the 
trilinear soft scalar coupling $\azero$ and $\sgn(\mu)$ -- the sign of
the supersymmetric Higgs/higgsino mass parameter $\mu$. For a fixed
value of $\tanb$, physical masses and couplings are obtained by
running various mass parameters, along with the gauge and Yukawa
couplings, from their common values at $\mgut$ down to $\mz$ by using
the renormalization group equations.  We compute the
mass spectra with version 2.2 of the package SUSPECT~\cite{suspect:ref}. 
\begin{figure}[t!]
\begin{center}
\begin{minipage}{16.0cm}
{\hspace*{-.5cm}\psfig{figure=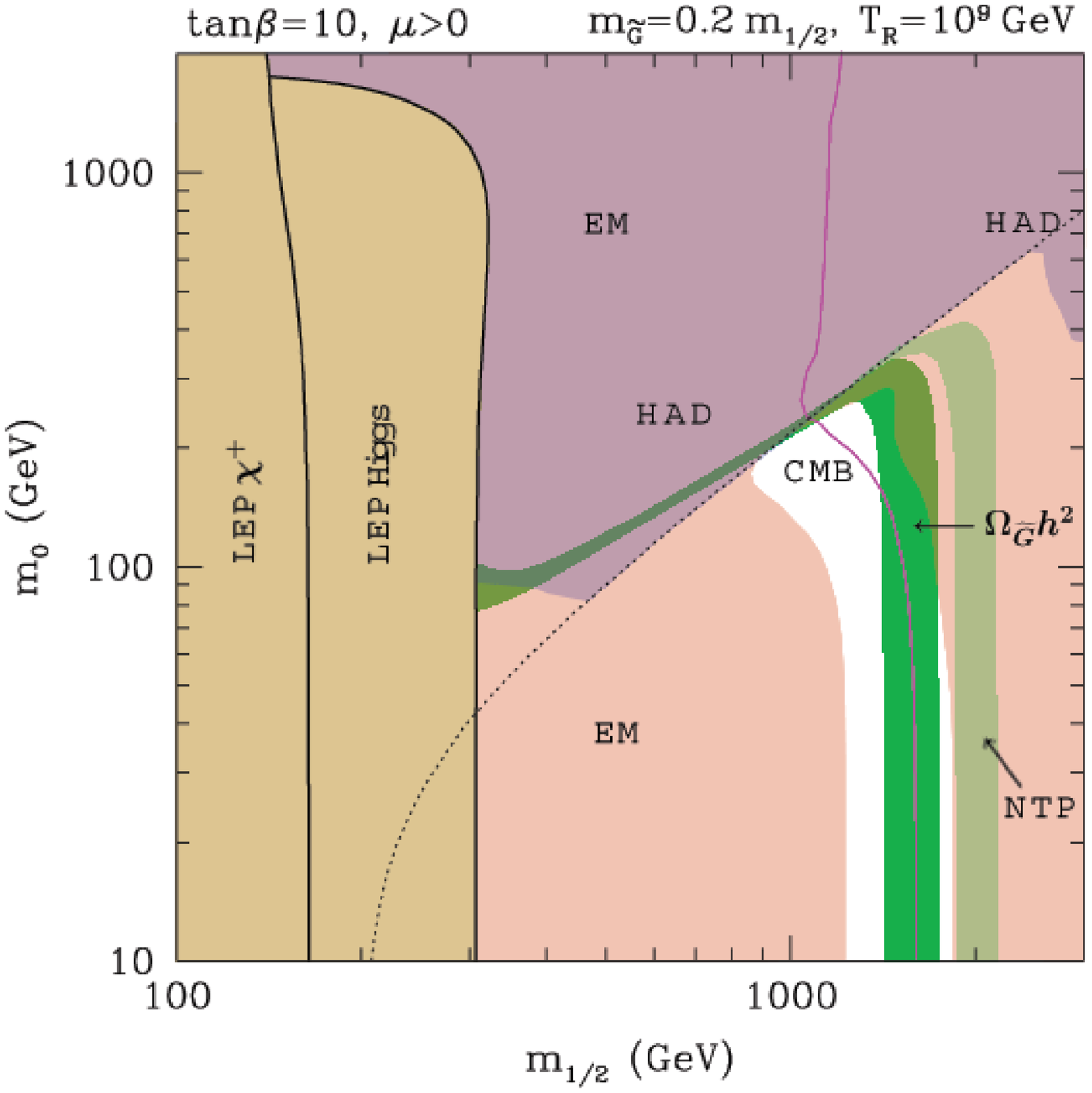, angle=0,width=8cm} 
 \hspace*{-.3cm}\psfig{figure=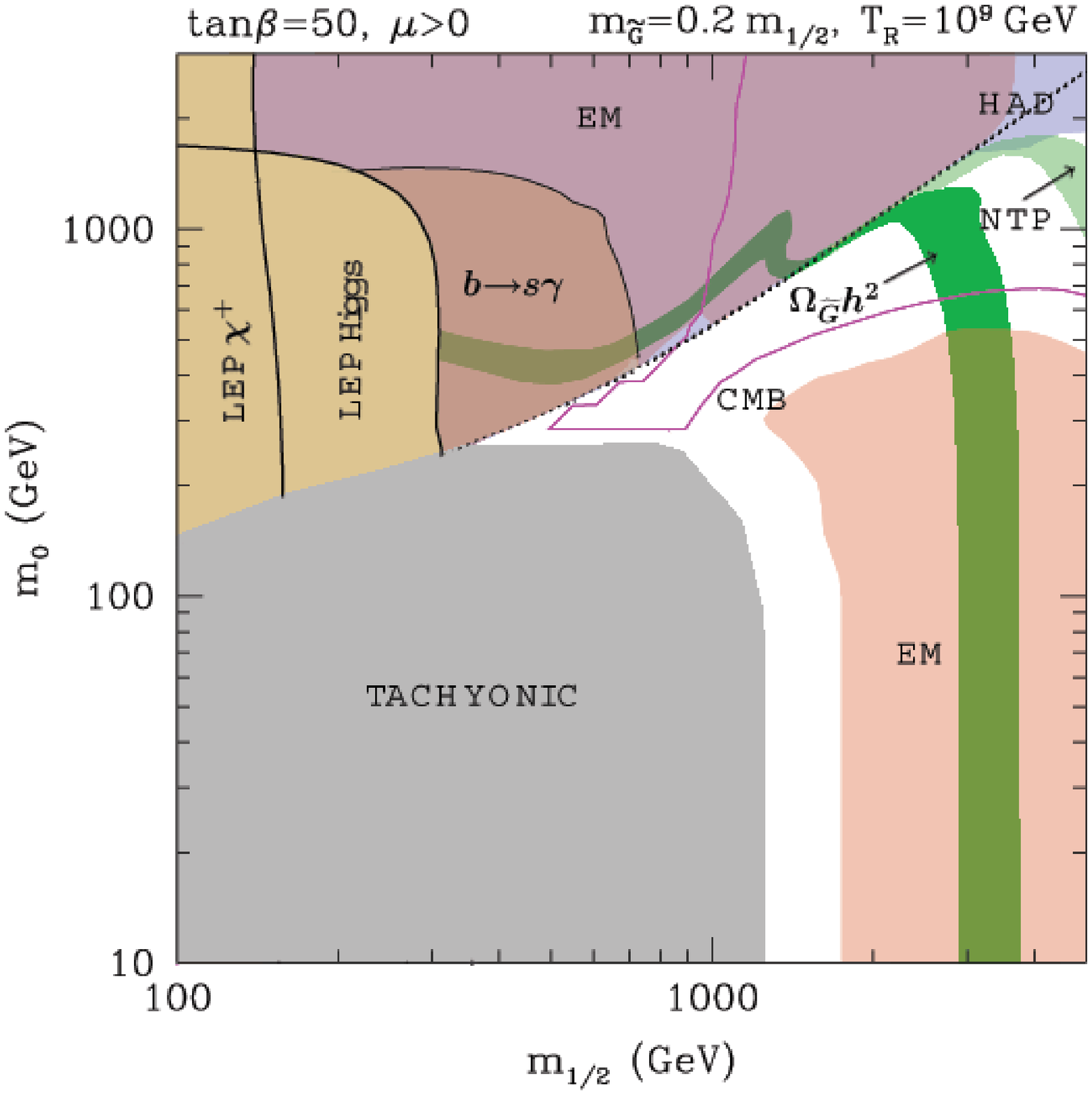, angle=0,width=8cm} 
}
\end{minipage}
\caption{\label{fig:tr1x9mg02mhalf} {\small The plane
($\mhalf,\mzero$) for $\tanb=10$ (left window) and $\tanb=50$ (right
window) and for $\azero=0$ and $\mu>0$. The light brown regions
labelled ``LEP $\chi^+$'' and ``LEP Higgs'' are excluded by
unsuccessful chargino and Higgs searches at LEP, respectively. In the
right window the darker brown regions labelled ``$b\to s\gamma$'' and
the dark grey region labelled ``TACHYONIC'' are also excluded. In the
dark green band labelled ``$\abundg$'' the total relic abundance of
the gravitino from both thermal and non--thermal production is in the
favored range, while in the light green regions which are denoted
``NTP'' the same is the case for the relic abundance from NTP
processes alone. Regions excluded by applying conservative bounds on
electromagnetic showers from $D/H+Y_p+{^7}{\! Li}$ obtained with
inputs~(\ref{cefodeutinput:eq})--(\ref{cefoliinput:eq}) are denoted in
orange and labelled ``EM''. Regions excluded by imposing less
conservative bounds on hadronic showers from $D/H+Y_p$ derived
assuming $B_{had}=1$ and
input~(\ref{kkmdeutinput:eq})--(\ref{kkmypinput:eq})
are denoted in blue and labelled ``HAD''. (The overlapping
EM/HAD--excluded regions appear as light violet.) A solid magenta
curve labelled ``CMB'' delineates the region (on the side of the
label) inconsistent with the CMB spectrum. The cosmologically favored
(green) regions are ruled out when we apply bounds from $D/H+Y_p$
derived with (\ref{kkmdeutinput:eq})--(\ref{kkmypinput:eq}) as input,
as described in the text.} }
\end{center}
\end{figure}

\begin{figure}[t!]
\begin{center}
\begin{minipage}{16.0cm}
{\hspace*{-.5cm}\psfig{figure=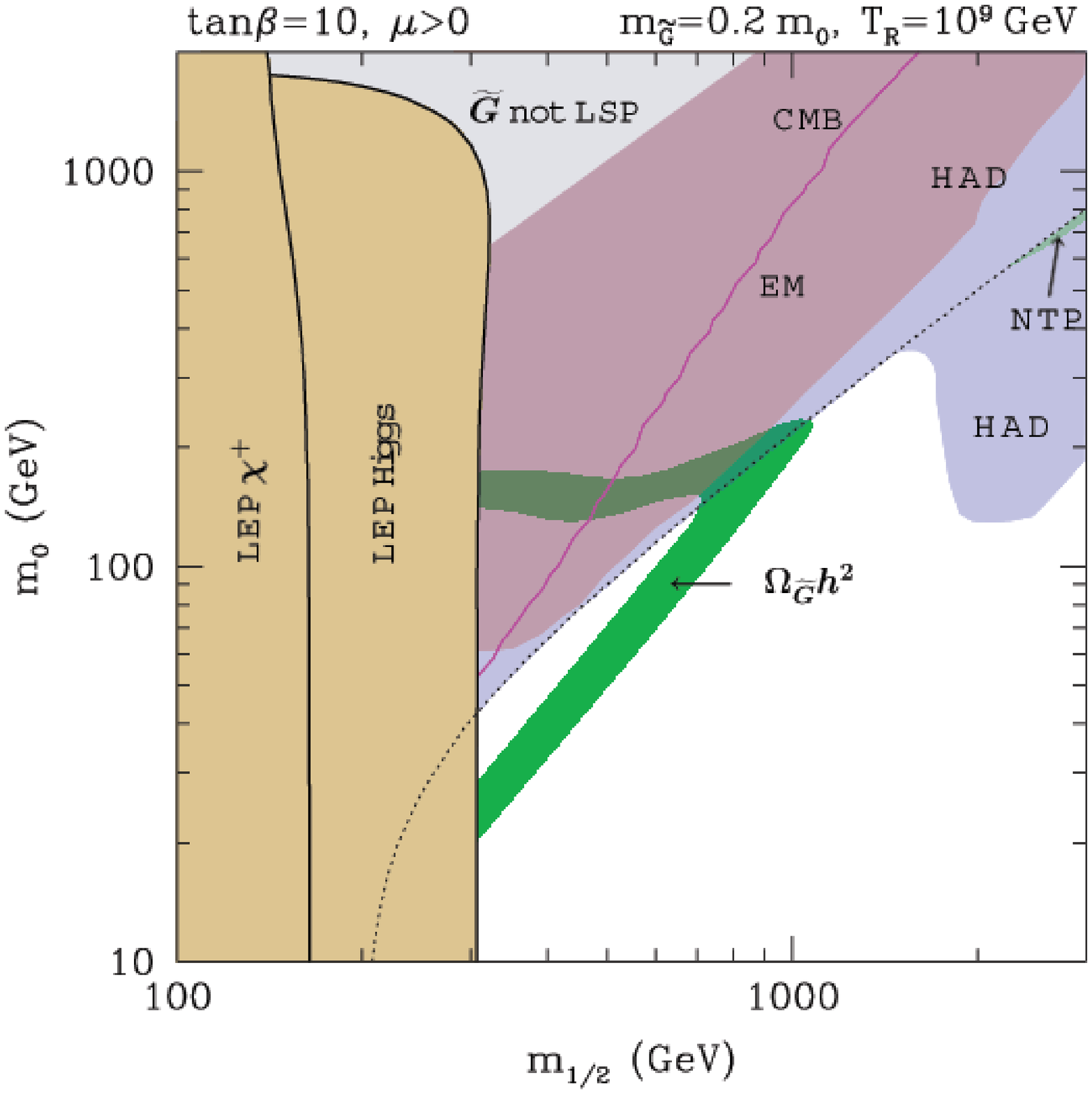, angle=0,width=8cm} 
 \hspace*{-.3cm}\psfig{figure=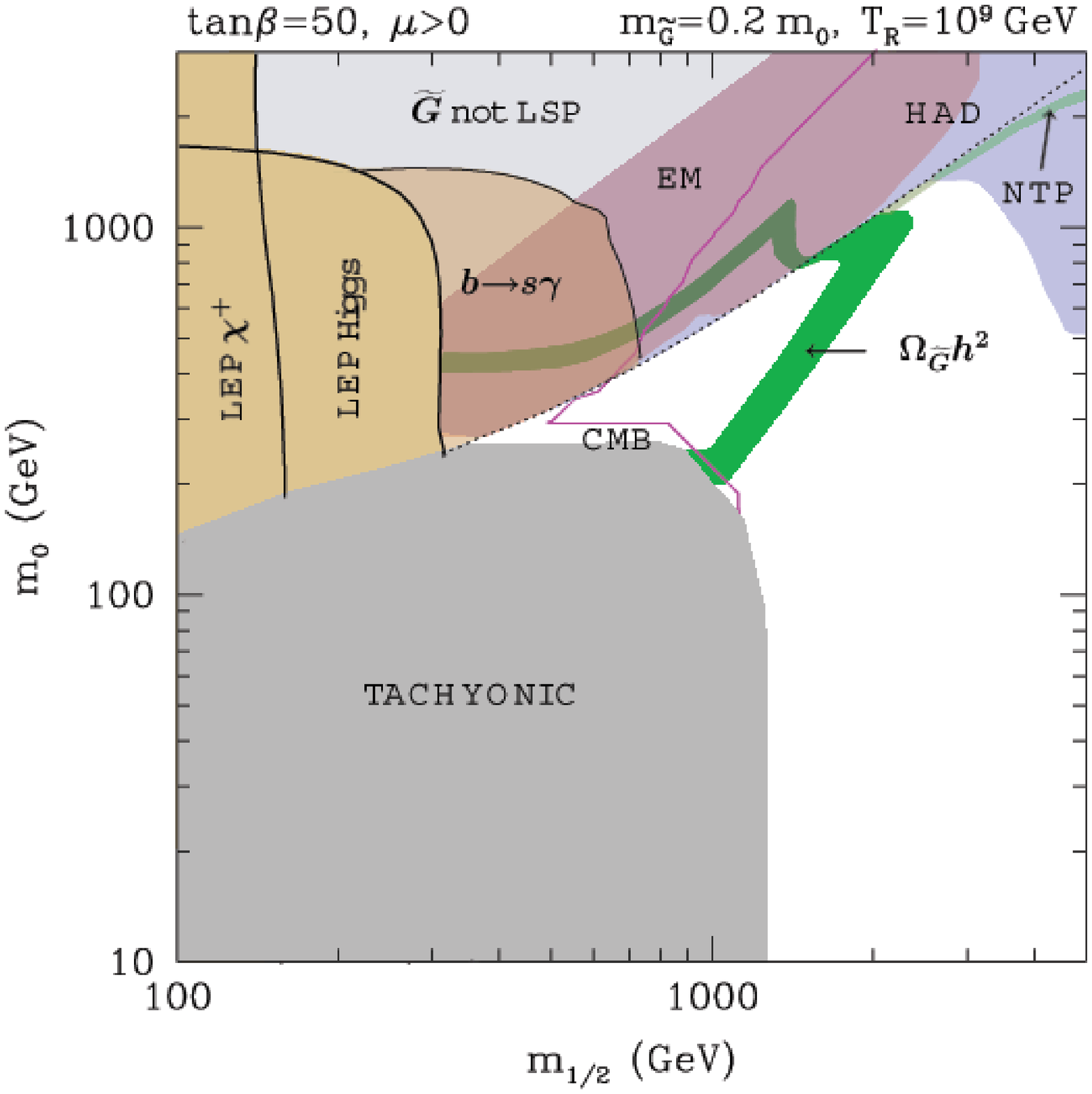, angle=0,width=8cm} 
}
\end{minipage}
\end{center}
\vspace*{-.35in} 
\hspace*{-.70in}
\begin{center}
\begin{minipage}{16.0cm}
{\hspace*{-.5cm}\psfig{figure=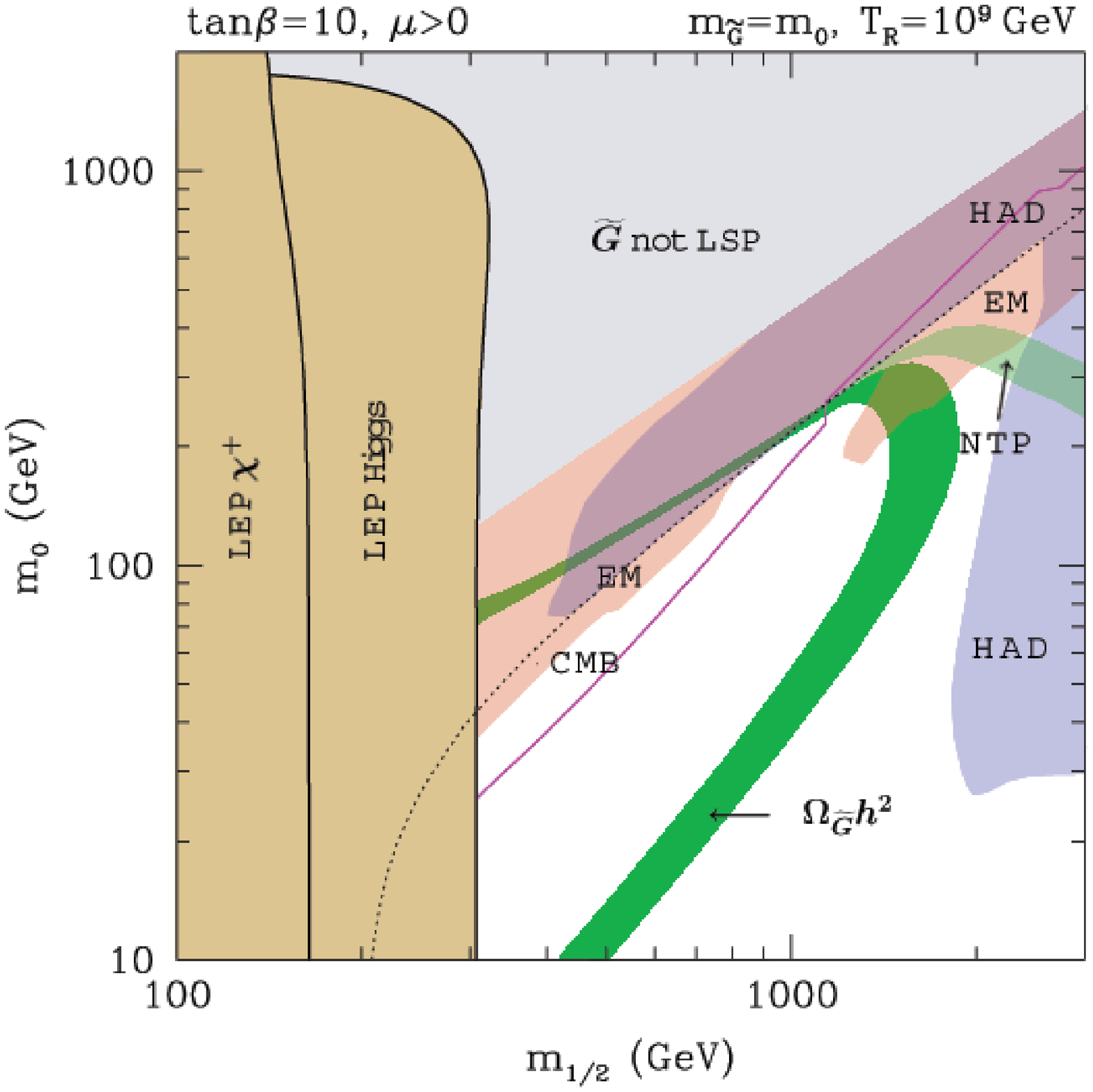, angle=0,width=8cm} 
 \hspace*{-.3cm}\psfig{figure=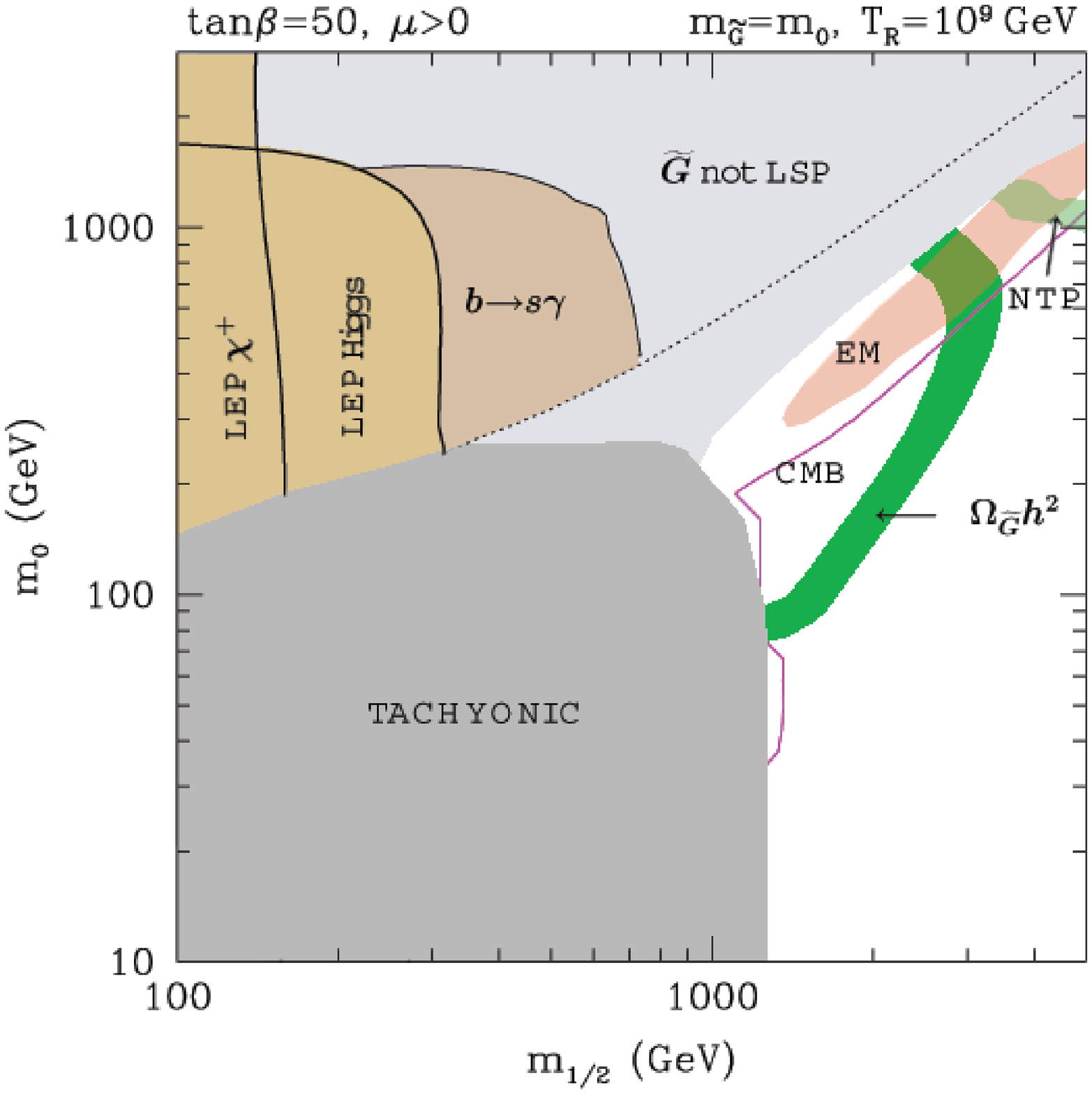, angle=0,width=8cm} 
}
\end{minipage}
\caption{\label{fig:tr1x9mgmzero} {\small The same as in 
fig.~\protect\ref{fig:tr1x9mg02mhalf} but for $\mgravitino=0.2\mzero$
(top row) and $\mgravitino=\mzero$ (bottom row). In the light grey
areas the gravitino is not the LSP. Applying bounds from $D/H+Y_p$ derived with
(\ref{kkmdeutinput:eq})--(\ref{kkmypinput:eq}) as input rules out most
of the cosmologically
favored regions, except for small patches for $\tanb=50$ in
the $\mgravitino=0.2\mzero$ and $\mzero$ windows, as described in the text.
 } }
\end{center}
\end{figure}

We present our results in the usual ($\mhalf,\mzero$) plane for two
representative choices of $\tanb=10$ and $50$ and for $\azero=0$ and
$\mu>0$. In light of the recent determinations from the Tevatron, we
fix the top mass at $m_t=178\gev$~\cite{tevatron_topmass}. 
In fig.~\ref{fig:tr1x9mg02mhalf} we consider the case
$\mgravitino=0.2\mhalf$ while in fig.~\ref{fig:tr1x9mgmzero} we take
$\mgravitino=0.2\mzero$ (top row) and $\mgravitino=\mzero$ (bottom row). In both
figures we fix  
the reheating temperature at $\treh=10^9\gev$.

To help understanding the figures, we remind the reader of some basic
mass relations. The mass of the gluino is roughly given by
$\mgluino\simeq 2.7\mhalf$.  The mass of the lightest neutralino,
which in the CMSSM is almost a pure bino, is
$\mchi\simeq0.4\mhalf$. The lighter stau $\stauone$ is dominated by
$\stauright$ and well above $\mz$ its mass is (neglecting Yukawa
contributions at large $\tanb$) roughly given by $\mstauone^2\simeq
\mzero^2+0.15\mhalf^2$.  This is why at $\mzero\ll\mhalf$ the stau is
lighter than the neutralino while in the other case the opposite is
true. The boundary between the two NLSP regions is marked in all the
figures with a roughly diagonal dotted line. (In the standard scenario
the region of a stable, electrically charged stau relic is thought to
be ruled out on astrophysics grounds.) Regions corresponding to the
lightest chargino and Higgs masses below their LEP
limits,~(\ref{eq:lepcharginobound}) and~(\ref{eq:lephiggsbound}),
respectively, are excluded. Separately marked for $\tanb=50$ is the
region inconsistent with the measured branching ratio $BR(B\rightarrow
X_s\gamma)$~(\ref{eq:bsgbounds}). (For $\tanb=10$, and generally not
too large $\tanb$, this constraint is much weaker and ``hidden''
underneath the above LEP bounds.) Here it is worth remembering that
the constraint is derived by assuming minimal flavor violation in the
squark sector -- the scenario where the mass mixing in the down--type
squark sector is the same as in the corresponding quark sector.
However, even small perturbation of the assumption may lead to a
significant relaxation (or strengthening) of the constraint from
$b\rightarrow s\gamma$~\cite{or1+2}. At small $\mzero$ and large
$\tanb$ some sfermions become tachyonic.  Finally, for some
combinations of parameters the gravitino is not the LSP. We exclude
such cases in this analysis.

Let us first concentrate on the regions where the total gravitino
relic abundance $\abundg$ is consistent with the preferred
range~(\ref{eq:abundgtot}). In all the windows these are represented
by green bands and labelled ``$\abundg$''. (On the left side of the
green bands $\abundg<0.094$ while on the other side $\abundg>0.129$.) Their shape
looks rather different in both the neutralino and the stau NLSP
regions. In the former, $\abundg$ is mostly determined by neutralino
decays (NTP mechanism), except that it is relaxed relative to the case
of neutralino LSP by the mass ratio $\mgravitino/\mchi$, as
in~(\ref{abundgntp:eq}). (Compare with the standard neutralino LSP
case in, \eg, fig.~5 of~\cite{crrs}.)  In this region of not too large
$\mhalf$ thermal production remains (so long as $\treh\lsim 10^9\gev$)
fairly inefficient, since it is proportional to $\mgluino$ which is
still not very large there (although growing with $\mhalf$).  In the right
column of the windows where $\tanb=50$, one can notice a characteristic
resonance due to efficient $\chi\chi$ annihilation via the
pseudoscalar Higgs $\ha$ exchange.

In all the windows light green bands (labelled ``NTP'') delineate the
regions below the dotted line (stau NLSP case) for which $\abundgntp$
is consistent with~(\ref{eq:abundgtot}). These regions become
cosmologically favored when one does not include TP or when $\treh\ll
10^9\gev$. Again, these are the regions where the relic abundance of
the stau (if it had been stable) would be consistent with the
range~(\ref{eq:abundgtot}), but shifted to the right and/or up by the
factor $\mgravitino/\mstauone$. 
Note that such regions correspond to ranges of $\mhalf$ beyond those
considered in~\cite{eoss03-grav} and were not identified there.

However, the contribution from thermal production, which linearly
grows with $\treh$~(\ref{eq:abundgbbb}), cannot be neglected when
$\treh$ is large.
In all the 
windows of figs.~\ref{fig:tr1x9mg02mhalf} and~\ref{fig:tr1x9mgmzero}
one can see how the effect of TP shifts the cosmologically preferred
region from the light green of $\abundgntp$ to the full green of
$\abundg$ where TP dominates. In fig.~\ref{fig:tr1x9mg02mhalf} both $\mgluino$ and
$\mgravitino$ scale up with $\mhalf$. As a result, TP dominated
regions of $\abundg$ are independent of $\mzero$
(compare~(\ref{eq:abundgbbb})). Even though in this figure the
contribution from TP is still subdominant, it does have a sizable
effect of shifting the vertical bands of total $\abundg$ to the left
of the ones due to NTP production alone.

In fig.~\ref{fig:tr1x9mgmzero}, $\abundgtp$ grows with $\mhalf$
(because $\mgluino\propto\mhalf$) but decreases with increasing
$\mzero$ (because $\mgravitino\propto\mzero$). Compare
again~(\ref{eq:abundgbbb}). This causes the green bands to bend
dramatically towards the diagonal.

Increasing $\mgravitino$ reduces the effect of TP. It simply becomes
harder to produce them in inelastic scatterings in the plasma. This
can be seen in fig.~\ref{fig:tr1x9mgmzero} by comparing the bottom row
(where $\mgravitino=\mzero$) with the top row (where
$\mgravitino=0.2\mzero$).  The green bands in the stau NLSP region,
where $\mgluino$ is large, are markedly shifted to the
right. (Increasing instead $\mgravitino$ as a fraction of $\mhalf$ is
a less promising way to go as this causes the region where the
gravitino is not the LSP to rapidly increase.)

It is thus clear that, so long as $\treh\leq10^9\gev$, one finds sizable
regions of rather large $\mhalf$ and much smaller $\mzero$ consistent
with the preferred range of CDM abundance. We now proceed to discussing
constraints from BBN and CMB.

Constraints from EM showers mainly due to hard photons in neutralino
NLSP decays into gravitinos~(\ref{gamchitogammagino:eq}) have
traditionally been regarded as severe, and this is confirmed in our
figures. Even with only the bounds from $D/H+Y_p+ {^7}{\! Li}$,
derived from conservative
inputs~(\ref{cefodeutinput:eq})--(\ref{cefoliinput:eq}), which are
labelled as ``EM'', most of the neutralino NLSP regions are ruled
out. One exception is when the number density of NLSP neutralinos is
reduced, like around the ``spike'' of the $\ha$ resonance at large
$\tanb=50$.  On the other hand, in the stau NLSP region the constraint
from EM showers is in most cases somewhat weaker.

Adding a constraint from ${^6}{\! Li}$ (not shown in the figures), as
in~\cite{eoss03-grav}, does not actually make much difference in all
the cases presented. Its main effect appears to be severely tempering,
to the point of almost removing, the spike  regions around the
$A$ resonance in the neutralino NLSP region.

On the other hand, adopting the sharper
inputs~(\ref{kkmdeutinput:eq})--(\ref{kkmypinput:eq}) into the bounds
from only $D/H+Y_p$ for constraining electromagnetic showers does
lead to a dramatic effect. 
We will discuss this case in more detail below.
 
Next we discuss an impact of the constraint from avoiding
excessive hadronic fluxes (labelled in the figures as ``HAD''). Applying
the bounds from only $D/H+Y_p$ but with less conservative
inputs~(\ref{kkmdeutinput:eq})--(\ref{kkmypinput:eq}) (and assuming
$B_{had}=1$),  basically rules out the whole neutralino NLSP
region, thus confirming the conclusions of Feng,
\etal,~\cite{fst04-sugra}. (The cases where the hadronic constraint is
stronger than the electromagnetic one are marked in blue. The opposite
case is marked in pink.) It also rules out some cases below the dotted
line where the very heavy stau NLSP decays fast enough to enhance the
importance of bounds from hadronic showers. It is, however, possible
that, with more conservative inputs, the hadronic constraint would not
be as severe even in the neutralino NLSP region. 

Last but not least, bounds on allowed distortions of the CMB spectrum
prove in many cases to be the most severe. They are shown as a solid
magenta lines. Regions on the side of the label ``CMB'' are ruled
out. While not as competitive in the neutralino NLSP region, for stau
NLSPs the CMB shape constraint due to the bound on the chemical
potential~(\ref{eq:mubound}), as already emphasized
in~\cite{fst04-sugra}, and also on the
$y$--parameter~(\ref{eq:yboundonxiem}), typically provide the tightest
constraint. The former bound, being applicable at not very late decay
times $\taux \gsim8.8\times10^{9}\sec$, excludes regions of the
$(\mhalf,\mzero)$ plane closer to the magenta curve, while the latter
bound, which applies to later decay times, excludes points at smaller
$\mhalf$ and/or $\mzero$.  It does not affect the magenta CMB
exclusion lines in the $(\mhalf,\mzero)$ plane.

\begin{figure}[t!]
\begin{center}
\begin{minipage}{16.0cm}
{\hspace*{-.5cm}\psfig{figure=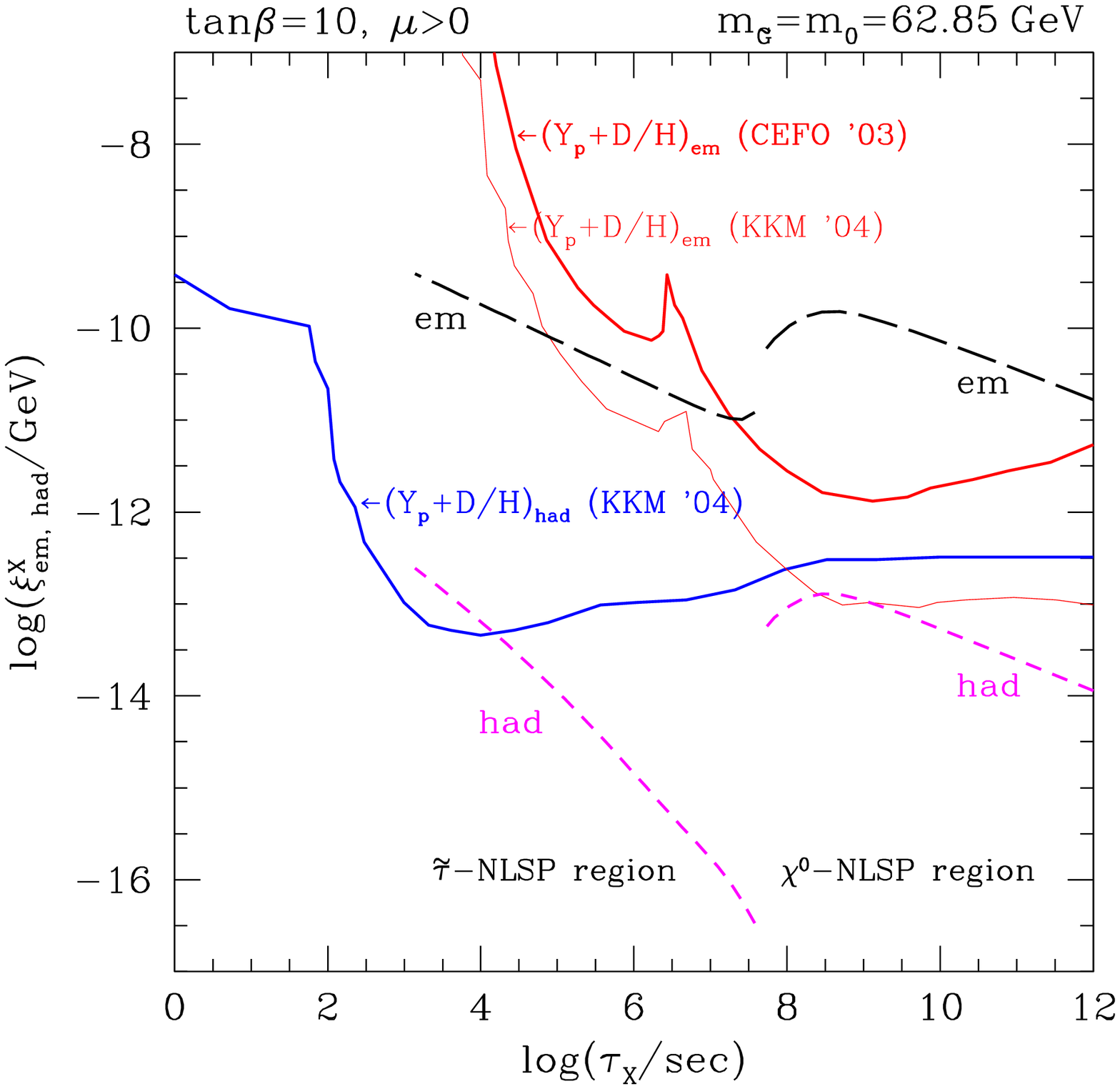, angle=0,width=8.0cm}
 \hspace*{-.3cm}\psfig{figure=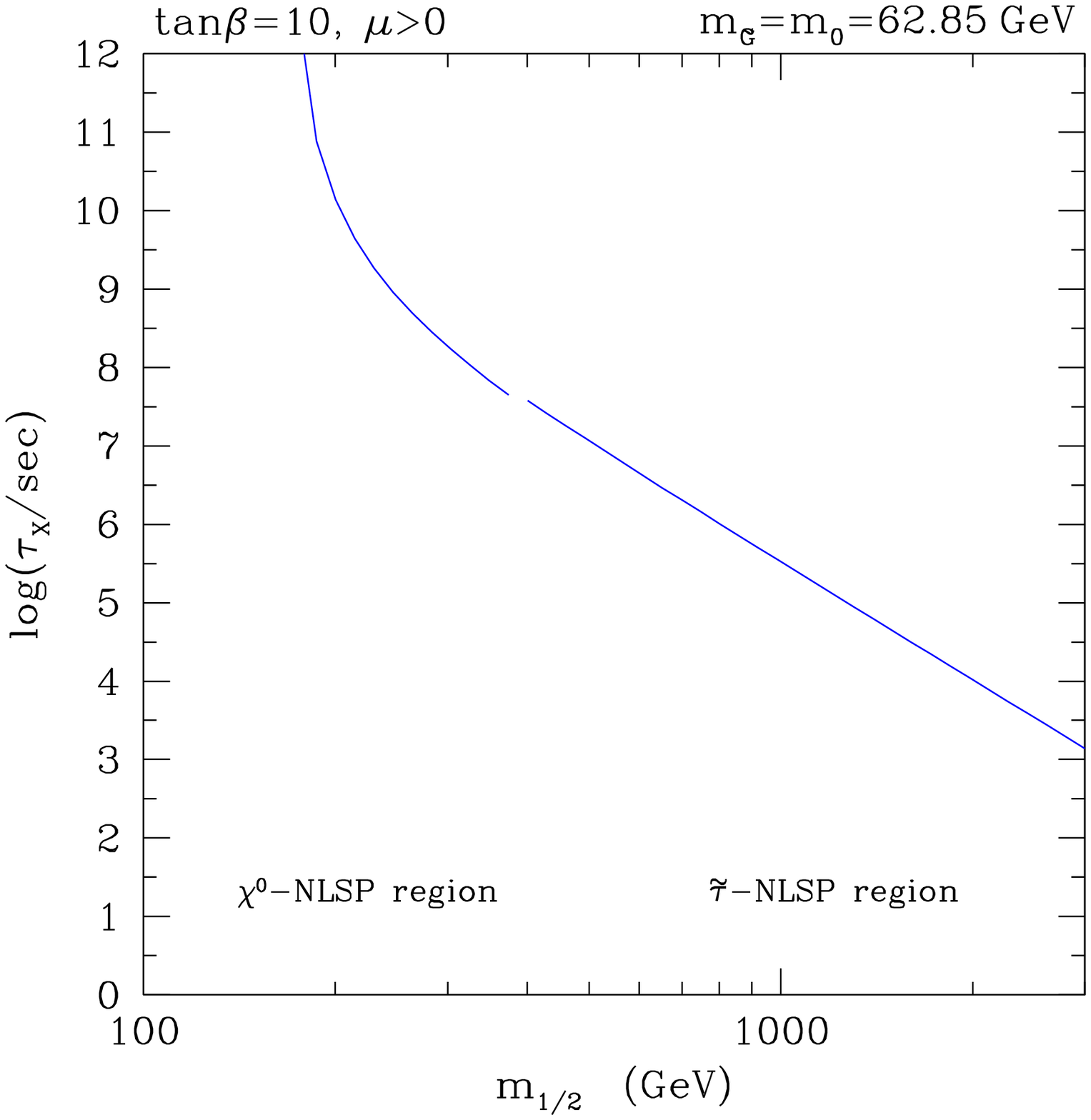, angle=0,width=8.0cm}
}
\end{minipage}
\caption{\label{fig:zetafigs} {\small Left panel: A comparison of the relative energy
 $\xi^X_{i}$ ($X=\chi,\stauone$ and $i={em},\, {had}$) with BBN
 constraints for the case
 $\tanb=10$, $\mu>0$ and $\mgravitino=\mzero=62.85\gev$. The black
 long--dashed curves corresponds to $\xi^X_{em}$ and should be
 compared with the red thick (thin) solid line denoting the upper
 bounds from $D/H+Y_p$ on the relative energy release from
 electromagnetic showers of Cyburt, \etal,~\cite{cefo02} (Kawasaki,
 \etal,~\cite{kkm04}), as explained in the text. Clearly, the excluded ranges
 of $\taux$ strongly depend on the assumed experimental
 bounds. The magenta short--dashed curves corresponds to $\xi^X_{had}$
 and should be compared with the  thick blue solid curve denoting the upper
 bounds from $D/H+Y_p$ on the relative energy release from hadronic
 showers only (no photo-dissociation) of~\cite{kkm04}, as explained in the text.
} }
\end{center}
\end{figure}

Coming back to the BBN constraints, in fig.~\ref{fig:zetafigs} we
illustrate a rather strong sensitivity of the excluded regions in the
$(\mhalf,\mzero)$ plane to the assumed abundances of the light
elements. We take $\tanb=10$, $\mgravitino=\mzero$ and fix
$\mzero=62.85\gev$.  In the left panel we plot the electromagnetic
energy release $\xi^X_{em}$ (black long--dashed line) and compare it
with the upper bounds taken from fig.~7 of~\cite{cefo02} (solid thick
red) and with the one corresponding to $D/H$ in fig.~42
of~\cite{kkm04} (solid thin red), as described in the previous
section. (Constraints from ${^7}{\! Li}$ are in this case weaker and
are not marked.) Clearly, while the $\chi$--NLSP region is excluded by
both upper bounds, much larger regions of $\taux$ are excluded in the
$\stauone$--NLSP region by the solid thin red line.  We can clearly
see that the excluded regions of the $(\mhalf,\mzero)$ plane strongly
depend on the assumed experimental bounds. When we apply the EM
constraints from~\cite{kkm04} to the cases presented here, in most of
them the cosmologically favored regions (both green bands of $\abundg$
and light green ones of $\abundgntp$) become excluded.

We also plot the hadronic energy release $\xi^X_{had}$ (magenta
short--dashed line) and compare it with the upper bounds taken from
fig.~43 of~\cite{kkm04} (solid thick blue line), as described in the
previous section. The right panel shows $\tau_X$ as a function of
$\mhalf$ in order to help relate fig.~\ref{fig:zetafigs} to the lower
left panel of fig.~\ref{fig:tr1x9mgmzero}.

\begin{figure}[t!]
\begin{center}
\begin{minipage}{16.0cm}
{\hspace*{-.5cm}\psfig{figure=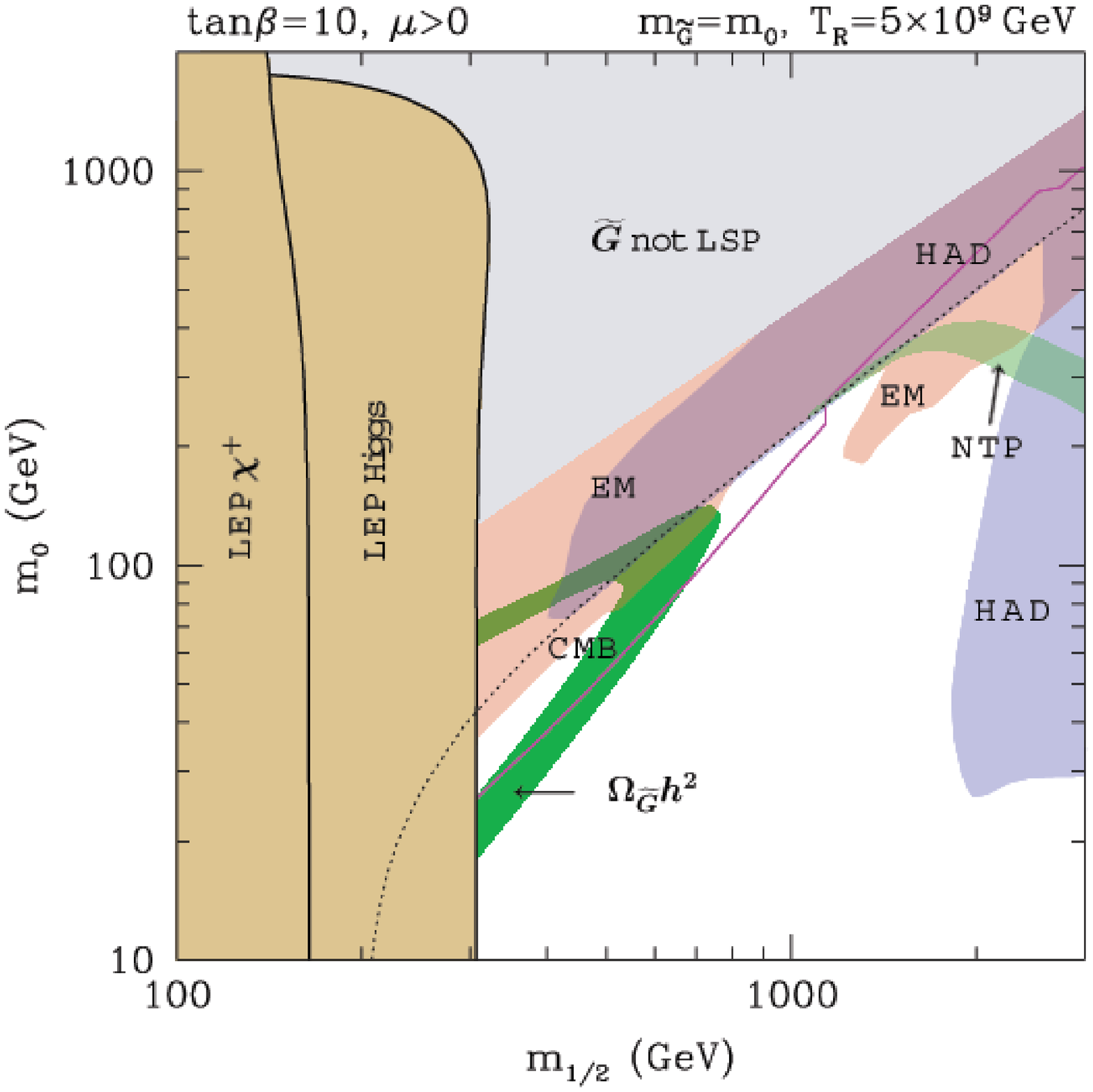, angle=0,width=8.0cm}
 \hspace*{-.3cm}\psfig{figure=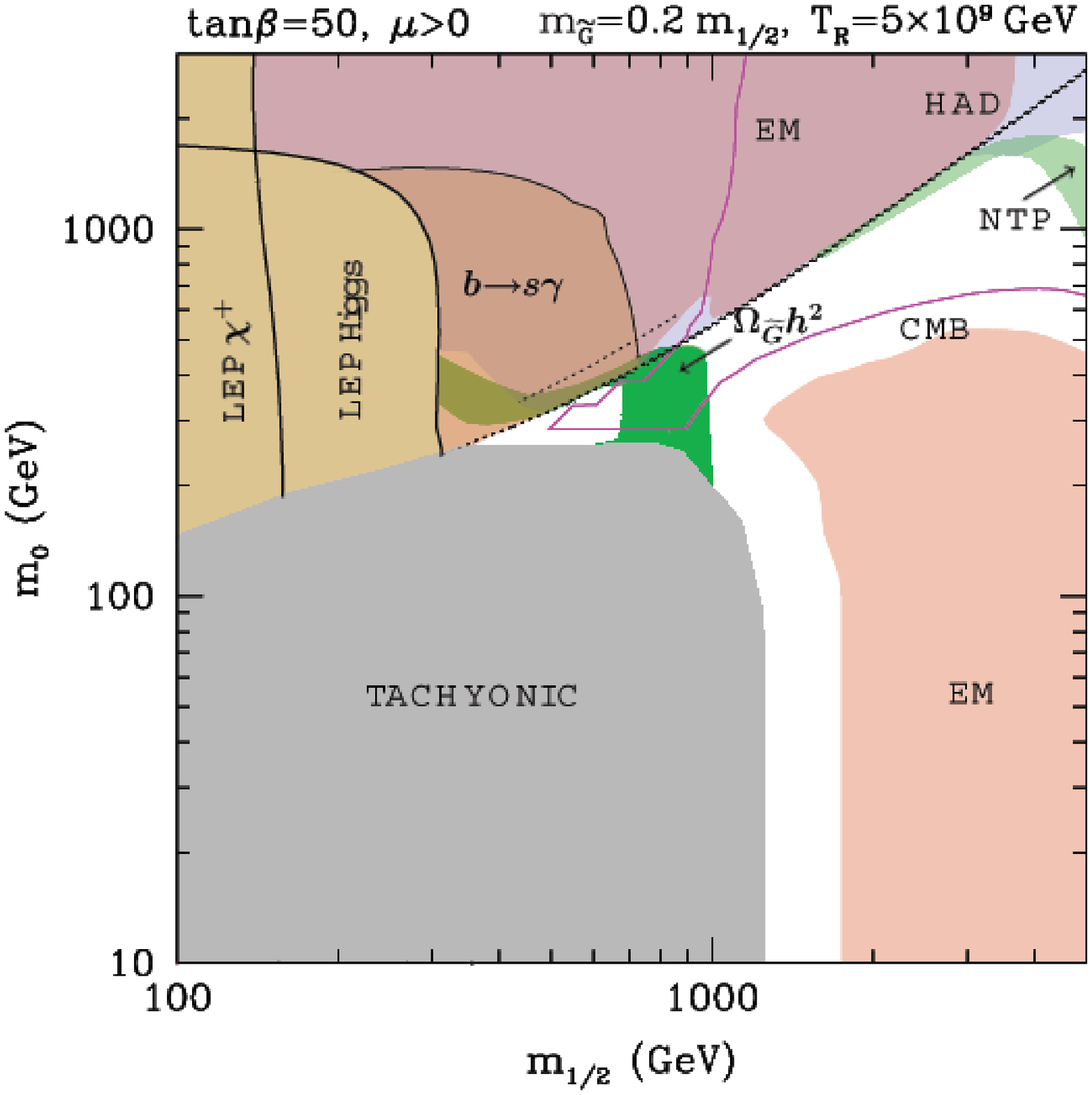, angle=0,width=8.0cm}
}
\end{minipage}
\caption{\label{fig:tr5x9} {\small The same as in 
fig.~\protect\ref{fig:tr1x9mg02mhalf} but for $\treh=5\times 10^9\gev$
and $\tanb=10$ and $\mgravitino=\mzero$
(left window) and $\tanb=10$ and $\mgravitino=0.2\mhalf$ (right
window). Applying bounds from $D/H+Y_p$ derived with
(\ref{kkmdeutinput:eq})--(\ref{kkmypinput:eq}) as input rules out the 
cosmologically favored regions, as described in the text.
 } }
\end{center}
\end{figure}

Given a significant squeeze on the gravitino CDM scenario imposed by
the interplay of the BBN and CMB constraints, a question arises
whether one can find allowed cases at $\treh$ even higher than
$10^9\gev$ presented in figs.~\ref{fig:tr1x9mg02mhalf}
and~\ref{fig:tr1x9mgmzero}.  As $\treh$ grows, the contribution from
TP also grows and the green band of the favored range of $\abundg$
moves left, towards excluded regions. In all the cases presented in
figs.~\ref{fig:tr1x9mg02mhalf} and~\ref{fig:tr1x9mgmzero}, except for
two, even a modest increase in $\treh$ is not allowed by our
conservative bounds from BBN and CMB. The two surviving cases are
presented in fig.~\ref{fig:tr5x9} for $\treh=5\times10^9\gev$. They
are already inconsistent with bounds on electromagnetic cascades from
$D/H+Y_p$ alone when one adopts the less conservative
inputs~(\ref{kkmdeutinput:eq})--(\ref{kkmypinput:eq}). The case in the
right window also becomes excluded by the bounds from $D/H+Y_p+
{^7}{\! Li}+ {^6}{\! Li}$, used in~\cite{eoss03-grav}, derived using
conservative inputs~(\ref{cefodeutinput:eq})--(\ref{cefoliinput:eq}).
Finally, an improvement of some order of magnitude in the upper
bound~(\ref{eq:mubound}) on $\mu$ would also rule these cases
out. Given the above discussion, it is unlikely that at higher $\treh$
any cases of the favored range of $\abundg$ will remain consistent
with CMB and/or BBN constraints.

It is interesting that, in all the cases presented above for which
$\mgravitino\propto\mzero$, the (pale green) region of
$\abundgntp\sim0.1$ is almost completely excluded by a combination of
BBN and CMB constraints. In other words, for this case, a contribution
to $\abundg$ from TP has to be substantial in order to escape the
above constraints. On the other hand, this is not appear to be the case for
$\mgravitino\propto\mhalf$. 

\section{Implications for Leptogenesis and SUSY Searches at the
  LHC}\label{sec:implications} 

It is interesting that, in spite of tight and improving constraints
from CMB and BBN, the possibility that in models of low energy SUSY
with gravity--mediated SUSY breaking the gravitino may be the main
component of the cold dark matter in the Universe remains
open. Clearly this is the case when the reheating temperature
$\treh\lsim10^{9}\gev$ for which a contribution from thermal
production can be neglected. Then the cosmologically favored regions
due to NLSP decays (non--thermal production) alone shift to larger
$\mhalf$, which is typically (albeit not always) somewhat less
affected by the above constraints.

Perhaps even more interesting is the fact that the reheating temperatures as
large as $5\times 10^9\gev$ seems to be allowed. This should be
encouraging to those who favor thermal leptogenesis as a way of
producing the baryon--antibaryon asymmetry in the Universe.

We stress that the above conclusions depend rather sensitively on the
assumed input from BBN bounds. Given a number of outstanding
discrepancies in determinations of the abundances of light elements,
especially of ${^6}{\! Li}$ and ${^3}{\! He}$, in our analysis we have
decided to 
apply rather conservative bounds derived using rather generous inputs, but
also discussed the (severe) impact of sharpening them.  In
constraining electromagnetic fluxes we included bounds from
$D/H+Y_p+{^7}{\! Li}$ derived with conservative 
inputs~(\ref{cefodeutinput:eq})--(\ref{cefoliinput:eq}). We did not take
into account bounds from ${^6}{\! Li}$ nor ${^3}{\! He}$ which otherwise
would be most constraining. 
The constraints on
hadronic fluxes that we have adopted are somewhat less conservative since they
are based on less generous
inputs~(\ref{kkmdeutinput:eq})--(\ref{kkmypinput:eq}) and on assuming
$B_{had}=1$.  
Improvements in determining 
the above ranges is likely to provide a very stringent
constraint on allowing high reheating temperatures 
$\treh\sim10^{9}\gev$, and perhaps even on the whole hypothesis of
gravitino CDM in the CMSSM and similar models.
Likewise, the (independent from BBN) bound from the CMB spectrum, if
improved by at least one order of magnitude, is likely to prove highly
constraining to the scenario. Certainly it would rule out the
presented above cases of $\treh=5\times 10^9\gev$.
Finally, we have neglected other possible, although strongly model
dependent, mechanisms of producing gravitinos, \eg\ from inflaton
decays, which could also contribute to $\abundgntp$.

An experimental verification of the scenario will, for the most part,
be rather challenging but not impossible. At the LHC one 
will have a good chance of exploring some interesting ranges of mass
parameters.
The most promising signature will be a detection of a massive, stable and
electrically charged (stau) particle.  In most cases, the
cosmologically favored regions correspond to $\mzero\lsim1\tev$ and
large $\mhalf\gsim1\tev$, up to $2\tev$ at small $\tanb$, and $4\tev$
at large $\tanb$. A significant fraction of this region should be
explored in the stau mode since $\mstauone^2\simeq
\mzero^2+0.15\mhalf^2$ and stable stau mass will probably be probed up
to $\sim1\tev$. Gluino search may be less promising since
$\mhalf\gsim1\tev$ implies $\mgluino\gsim2.7\tev$, which will be just
outside of the reach of the LHC. However, some interesting cases may
still be explored. One is that of fairly small $\tanb$ and
$\mgravitino\propto\mzero$, as in the left windows of
figs.~\ref{fig:tr1x9mgmzero} and~\ref{fig:tr5x9}. The other is that of
large $\tanb$ and large $\treh$, as in the right window of
fig.~\ref{fig:tr5x9}. On the other hand, since in the cosmologically
favored regions $\mzero\ll\mhalf$ squarks (and sleptons) will be
fairly light and lighter than the gluino. More work will be needed to
more fully assess to what extend the cosmologically favored regions
will be explored at the LHC. Finally, the staus will eventually
decay. The proposal of~\cite{bhrey04} of measuring their delayed
decays is in this context worth pursuing as it would give a unique
opportunity of experimentally exploring the hypothesis of gravitino
cold dark matter and of probing the Planck scale at the LHC.

\acknowledgments
LR would like to thank L.~Covi, K.~Jedamzik, C.~Mu{\~n}oz and H.-P. Nilles
for helpful comments and the CERN Physics Department, Theory Division where the
work has been completed, for its kind hospitality. We also acknowledge
the funding from EU FP6 programme - ILIAS (ENTApP).\\


\begin{thebibliography}{99}

\bibitem{dz81}
A.D.~Dolgov and Ya.B.~Zel{\'{ }}dovich, \rmp{53}{1981}{1}.

\bibitem{pp82}
H.~Pagels and J.R.~Primack,
\prl{48}{1982}{223}.

\bibitem{weinberg-grav82}
S.~Weinberg,
\prl{48}{1982}{1303}.

\bibitem{khlopov+line84}
M.Y.~Khlopov and A.~Linde,
\plb{138}{1984}{265}.

\bibitem{ekn84}
J.R.~Ellis, J.E.~Kim and D.V.~Nanopoulos,
\plb{145}{1984}{181}.

\bibitem{ens84}
J.R.~Ellis, D.V.~Nanopoulos and S.~Sarkar, 
\npb{259}{1985}{175}.

\bibitem{nos83}
D.V.~Nanopoulos, K.A.~Olive and M.~Srednicki, 
\plb{127}{1983}{30}.

\bibitem{jss85}
R.~Juszkiewicz, J.~Silk and A.~Stebbins,
\plb{158}{1985}{463}.

\bibitem{km94}
M.~Kawasaki and T.~Moroi,
\ptp{93}{1995}{879} [arXiv:hep-ph/9403364].

\bibitem{cefo02}
R.H.~Cyburt, J.~Ellis, B.D.~Fields and K.A.~Olive, 
\prd{67}{2003}{103521} [arXiv:astro-ph/0211258].

\bibitem{kkm04} 
M.~Kawasaki, K.~Kohri and T.~Moroi, arXiv:astro-ph/0402490 and
arXiv:astro-ph/0408426.

\bibitem{mmy93}
T.~Moroi, H.~Murayama and M.~Yamaguchi,
\plb{303}{1993}{289}.

\bibitem{pq}
R.~D.~Peccei and H.~R.~Quinn, \prl{38}{1977}{1440} and \prd{16}{1977}{1791}.

\bibitem{kmn}
J.~E.~Kim, A.~Masiero and D.~V.~Nanopoulos, \plb{139}{1984}{346}.    
 
\bibitem{bgm}
S.~A.~Bonometto, F.~Gabbiani and A.~Masiero,
\plb{222}{1989}{433} and \prd{49}{1994}{3918}
[arXiv:hep-ph/9305237].

\bibitem{rtw}
K.~Rajagopal, M.~S.~Turner and F.~Wilczek, 
\npb{358}{1991}{447}.

\bibitem{ckkr}
L.~Covi, H.~B.~Kim, J.~E.~Kim and L.~Roszkowski, 
\jhep{05}{2001}{033} [arXiv:hep-ph/0101009].

\bibitem{ay00}
T.~Asaka and T.~Yanagida, 
\plb{494}{2000}{297} [arXiv:hep-ph/0006211].

\bibitem{ckr}
L.~Covi, J.~E.~Kim and L.~Roszkowski, 
\prl{82}{1999}{4180} [arXiv:hep-ph/9905212].

\bibitem{crs1} 
L.~Covi, L.~Roszkowski and M.~Small, 
\jhep{07}{2002}{023} [arXiv:hep-ph/0206119].

\bibitem{bs04}
A.~Brandenburg and F.D.~Steffen, 
\jcap{08}{008}{2004}  [arXiv:hep-ph/0405158].

\bibitem{bbp98} 
M.~Bolz, W.~Buchm\"uller and Pl\"umacher,
\plb{443}{1998}{209} [arXiv:hep-ph/9809381].

\bibitem{bbb00} 
M.~Bolz, A.~Brandenburg and W.~Buchm\"uller, 
\npb{606}{2001}{518} [arXiv:hep-ph/0012052].

\bibitem{fy}
M.~Fukugita and T.~Yanagida,
\plb{174}{1986}{45}.

\bibitem{crv96}
L.~Covi, E.~Roulet and F.~Vissani,
\prd{384}{1996}{169} [arXiv:hep-ph/9605319].

\bibitem{gnrrs04} 
G.F.~Giudice, A.~Notari, M.~Raidal, A.~Riotto and A.~Strumia, 
\npb{685}{2004}{89} [arXiv:hep-ph/031012].

\bibitem{feng03-prl}
J.L.~Feng, A.~Rajaraman and F.~Takayama, 
\prl{91}{2003}{011302} [arXiv:hep-ph/0302215].

\bibitem{feng03} 
J.L.~Feng, A.~Rajaraman and F.~Takayama,
\prd{68}{2003}{063504} [arXiv:hep-ph/0306024].

\bibitem{fst04-slep} 
J.L.~Feng, S.~Su and  F.~Takayama, arXiv:hep-ph/0404198.

\bibitem{fst04-sugra} 
J.L.~Feng, S.~Su and  F.~Takayama, arXiv:hep-ph/0404231.

\bibitem{eoss03-grav}
J.~Ellis, K.A.~Olive, Y.~Santoso and V.~Spanos, 
\plb{588}{2004}{7} [arXiv:hep-ph/0312262].

\bibitem{kkrw94}
G.~L.~Kane, C.~Kolda, L.~Roszkowski, and J.~D.~Wells, 
\prd{49}{1994}{6173} [arXiv:hep-ph/9312272].

\bibitem{ahs00}
T.~Asaka, K.~Hamaguchi and K.~Suzuki,
\plb{490}{2000}{136} [arXiv:hep-ph/0005136].

\bibitem{ggr98}
T.~Gherghetta, G.F.~Giudice and A.~Riotto, 
\plb{446}{1999}{28} [arXiv:hep-ph/9808401].

\bibitem{grt99}
G.F.~Giudice, A.~Riotto and I.~Tkachev,
\jhep{9911}{1999}{036}  [arXiv:hep-ph/9911302] and
\jhep{9908}{1999}{009} [arXiv:hep-ph/9907510].
     
\bibitem{kklp99}
R.~Kallosh, L.~Kofman, A.~Linde and A.~Van Proeyen,
\prd{61}{2000}{103503} [arXiv:hep-th/9907124].

\bibitem{nps01}
H.P.~Nilles, M.~Peloso and L.~Sorbo, 
\prl{87}{2001}{051302} [arXiv:hep-ph/0102264].

\bibitem{kyy04} For a recent analysis, see
K.~Kohri, M.~Yamaguchi and J.~Yokoyama, arXiv:hep-ph/0403043.

\bibitem{hw04}
D.~Hooper and L.-T.~Wang, 
arXiv:hep-ph/0402220.

\bibitem{kt90}
E.W. Kolb and M.S. Turner, 
{\em The Early Universe}, Addison-Wesley (1990).

\bibitem{moroi95}
T.~Moroi, PhD thesis,  arXiv:hep-ph/9503210.

\bibitem{nrr1}
T.~Nihei, L.~Roszkowski, R.~Ruiz de Austri, 
\jhep{05}{2001}{063} [arXiv:hep-ph/0102308].

\bibitem{nrr2}
T.~Nihei, L.~Roszkowski, R.~Ruiz de Austri,
\jhep{03}{2002}{031} [arXiv:hep-ph/0202009].

\bibitem{gondoloedsjo}
J.~Edj\"o and P.~Gondolo, 
\prd{56}{1997}{1879} [arXiv:hep-ph/9704361].

\bibitem{nrr3}
T.~Nihei, L.~Roszkowski, R.~Ruiz de Austri,
\jhep{07}{2002}{024} [arXiv:hep-ph/0206266].

\bibitem{rrn1} 
L.~Roszkowski, R.~Ruiz de Austri and T.~Nihei,
\jhep{08}{2001}{024} [arXiv:hep-ph/0106334].
%

\bibitem{crrs}
L.~Covi, L.~Roszkowski, R.~Ruiz de Austri and M.~Small, 
\jhep{0406}{2004}{003} [arXiv:hep-ph/0402240].


\bibitem{renoseckel88}
M.H.~Reno and D.~Seckel,
\prd{37}{1988}{3441}. 

\bibitem{dehs88+89}
S.~Dimopoulos, R.~Esmailzadeh, L.J.~Hall and G.D.~Starkman,
\apj{330}{1988}{545} and
\npb{311}{1989}{699}.

\bibitem{jedamzik04}
K.~Jedamzik, 
\prd{70}{2004}{063524} [arXiv:astro-ph/0402344].

\bibitem{jedamzik99}
K.~Jedamzik, 
\prl{84}{2000}{3248} [arXiv:astro-ph/9909445].

\bibitem{sjsb95}
G.~Sigl, K.~Jedamzik, D.N.~Schramm, V.S.~Berezinsky, 
\prd{52}{1995}{6682} [arXiv:astro-ph/9503094].

\bibitem{wmap_cdm}
D.~N.~Spergel, {\it et~al.}, 
\apj{148}{2003}{175} [arXiv:astro-ph/0302209].

\bibitem{wmap_eta}
C.~L.~Bennett, \etal,
\apj{148}{2003}{1} [arXiv:astro-ph/0302207].

\bibitem{mubound}
D.~J.~Fixsen, \etal,
\apj{473}{1996}{576}
[astro-ph/9605054];
K.~Hagiwara, \etal,  [Particle Data Group],
\prd{66}{2002}{010001}.

\bibitem{hu93} W.~Hu and J.~Silk, 
\prl{70}{1993}{2661} and 
\prd{48}{1993}{485}.

\bibitem{pdg02} K. Hagiwara, \etal, 
\prd{66}{2002}{010001}.


\bibitem{lepbounds}
LEPSUSYWG, ALEPH, DELPHI, L3 and OPAL experiments, note LEPSUSYWG/01-03.1
(http://lepsusy.web.cern.ch/lepsusy/Welcome.html). 

\bibitem{or1+2}
K.~Okumura and L.~Roszkowski, 
\prl{92}{2004}{161801} [arXiv:hep-ph/0208101] 
and 
\jhep{0310}{2003}{024} [arXiv:hep-ph/0308102].

\bibitem{suspect:ref}
A.~Djouadi, J.-L.~Kneur and G.~Moultaka. The package 
SUSPECT is available at 
http://www.lpm.univ-montp2.fr:7082/{\~{ }}kneur/suspect.html.

\bibitem{tevatron_topmass} See, \eg, L.~Cerrito, arXiv:hep-ex/0405046.

\bibitem{bhrey04} 
W.~Buchm\"uller, K.~Hamaguchi, M.~Ratz and T.~Yanagida,
\plb{588}{2004}{90} [arXiv:hep-ph/0402179].

\end{thebibliography}
\end{document}